\documentclass[structabstract]{aa}  
\usepackage{graphicx}
\begin{document}

\input{psfig}



\newbox\grsign \setbox\grsign=\hbox{$>$} \newdimen\grdimen \grdimen=\ht\grsign
\newbox\simlessbox \newbox\simgreatbox
\setbox\simgreatbox=\hbox{\raise.5ex\hbox{$>$}\llap
     {\lower.5ex\hbox{$\sim$}}}\ht1=\grdimen\dp1=0pt
\setbox\simlessbox=\hbox{\raise.5ex\hbox{$<$}\llap
     {\lower.5ex\hbox{$\sim$}}}\ht2=\grdimen\dp2=0pt
\def\simgreat{\mathrel{\copy\simgreatbox}}
\def\simless{\mathrel{\copy\simlessbox}}
\newbox\simppropto
\setbox\simppropto=\hbox{\raise.5ex\hbox{$\sim$}\llap
     {\lower.5ex\hbox{$\propto$}}}\ht2=\grdimen\dp2=0pt
\def\simpropto{\mathrel{\copy\simppropto}}

\title{ Zinc abundances in Galactic bulge field red giants: Implications for DLA systems
\thanks{Observations collected both  at the European  Southern  Observatory,
  Paranal,  Chile  (ESO programmes  71.B-0617A, 73.B0074A, and GTO 71.B-0196)} }
\author{
B. Barbuy\inst{1}
\and
A. C. S. Fria\c ca\inst{1}
\and
C. R. da Silveira\inst{1}
\and
V. Hill\inst {2}
\and
M. Zoccali\inst{3,4}
\and
D. Minniti\inst{4,5}
\and
A. Renzini\inst{6}
\and
S. Ortolani\inst{7}
\and
A. G\'omez\inst{8}
}
\offprints{B. Barbuy}
\institute{
Universidade de S\~ao Paulo, IAG, Rua do Mat\~ao 1226,
Cidade Universit\'aria, S\~ao Paulo 05508-900, Brazil\\
 e-mail: barbuy@astro.iag.usp.br
\and
Universit\'e de Sophia-Antipolis,
 Observatoire de la C\^ote d'Azur, CNRS UMR 6202, BP4229, 
06304 Nice Cedex 4, France
\and
Universidad Catolica de Chile, Instituto de Astrofisica,
Casilla 306, Santiago 22, Chile
\and
Millenium Institute of Astrophysics, Av. Vicu\~na Mackenna 4860,
Macul, Santiago, Chile
\and
Departamento de Ciencias Fisicas, Universidad Andres Bello, Republica 220, Santiago, Chile
\and
Osservatorio Astronomico di Padova, Vicolo
 dell'Osservatorio 5, I-35122 Padova, Italy
\and
Universit\`a di Padova, Dipartimento di Astronomia, Vicolo
 dell'Osservatorio 2, I-35122 Padova, Italy
\and
Observatoire de Paris-Meudon,  92195 Meudon Cedex, France
}

   \date{}

 
  \abstract
   {Zinc in stars is an important reference element because 
it is a proxy to Fe in studies of damped Lyman-$\alpha$ systems, 
permitting a comparison of chemical evolution histories of bulge
 stellar populations and DLAs.
In terms of nucleosynthesis, it behaves as an alpha element because it is enhanced
in metal-poor stars. Abundance studies in different stellar populations can give hints to 
the Zn production in different sites. }
   {The aim of this work is to derive the iron-peak element Zn 
abundances in 56 bulge giants from high resolution spectra. These results
are compared with data from other bulge samples, as well as from disk and halo stars, and damped Lyman-$\alpha$ systems, in order to better understand
 the chemical evolution in these environments. }
   {High-resolution spectra were obtained using  FLAMES+UVES on the 
Very Large Telescope. We computed the Zn abundances using the 
\ion{Zn}{I} lines at 4810.53 and 6362.34 {\rm \AA}. 
We considered the strong depression
in the continuum of the  \ion{Zn}{I} 6362.34 {\rm \AA} line, which is 
caused by the wings of the 
\ion{Ca}{I} 6361.79 {\rm \AA}  line suffering from autoionization.
CN lines blending the  \ion{Zn}{I} 6362.34 {\rm \AA} line are also included
in the calculations.
   }
   {We find [Zn/Fe]=+0.24$\pm$0.02 in the range $-$1.3 $<$ [Fe/H] $<$ $-$0.5
and [Zn/Fe]=+0.06$\pm$0.02 in the range $-$0.5 $<$ [Fe/H] $<$ $-$0.1,
whereas for [Fe/H]$\geq$ $-$0.1, it shows a spread of 
$-$0.60 $<$ [Zn/Fe] $<$ +0.15,
with most of these stars having low [Zn/Fe]$<$0.0. 
These low zinc abundances at the high metallicity end of the bulge
 define a decreasing trend in [Zn/Fe] with increasing metallicities.
  A comparison with Zn abundances in DLA systems is presented, 
where a dust-depletion
correction was applied for both Zn and Fe. 
When we take these corrections into account,
the [Zn/Fe] vs. [Fe/H] of the DLAs fall in the same region as the 
thick disk and bulge stars.
Finally, we  present a chemical evolution model of Zn enrichment in 
massive spheroids, representing a typical classical bulge evolution.}
   {}


   \maketitle
%

\section{Introduction} 

Zinc is an interesting element to study 
 because it can be observed in 
damped Lyman-$\alpha$ systems (DLAs), where it is assumed
as a proxy for Fe and where it provides most of our knowledge
of the chemical evolution of the Universe at high redshift,
through abundances in DLAs (Pettini et al. 1999; Prochaska 
\& Wolfe 2002). 

Zinc abundance derivation is important given its production in
different nucleosynthesis processes and environments:
weak s-process in hydrostatic phases of He and C burning in massive stars, 
 complete and incomplete Si-burning, explosive burning in core-collapse SNe,
and the main s-process in low and intermediate mass stars
 (e.g. Umeda \& Nomoto 2002; Bisterzo et al. 2004).

Zinc is in the so-called upper iron group with
atomic masses in the range 57$\leq$A$\leq$ 66,
which includes species up to $^{66}$Zn (Woosley \& Weaver 1995). 
Umeda \& Nomoto (2002) show that in massive stars,
 the iron-peak elements Cr, Mn, Co, and Zn
are produced in complete Si-burning with a peak temperature 
T$_{\rm peak}$ $>$ 5$\times$10$^{9}$ K and 
in incomplete Si-burning at temperatures in the range
 4$\times$10$^{9}$ $<$ T$_{\rm peak}$ $<$ 5$\times$10$^{9}$ K. 
They also found that [Zn/Fe] is greater for deeper mass cuts
in the explosion process,
 smaller neutron excess, and higher explosion energies
and that a higher Zn abundance results from deep mixing 
of complete Si-burning products and a fallback. 
At lower metallicities, $^{64}$Zn is produced
in complete Si-burning (cf. Umeda \& Nomoto 2003, 2005).
In addition, hypernovae, defined as supernovae with high explosion 
energies
 (E$_{51}$ $\simgreat$ 2 for M$\sim$13 M$_{\odot}$ 
and E$_{51}$ $\simgreat$ 20 for M $\simgreat$ 20 M$_{\odot}$), 
give rise to ejecta with [Zn/Fe] as high as $\sim$0.5 (Umeda \& Nomoto 2002;
Nomoto et al. 2013).
 Therefore, Kobayashi et al. (2006) have invoked hypernovae to explain
 the high [Zn/Fe] ratios in metal-poor stars.


The majority of the Fe-peak elements show solar abundance ratios in most 
objects for all metallicities. The elements Sc, Mn, Cu, and Zn, however, show different
trends (e.g. Sneden et al. 1991; Nissen et al. 2000; Ishigaki et al. 2013;
Barbuy et al. 2013).
Nissen \& Schuster (2011) find that Zn behaves like alpha elements,
with high-alpha halo stars and thick disk
stars  displaying also high Zn abundances,  whereas the
low-alpha halo stars show a lower Zn enhancement that decreases with metallicity. 
This distinct behaviour is explained if both these elements (Zn and alphas) are produced by core-collapse supernovae. The [Zn/Fe] decrease with metallicity in low-alpha stars is then expected, characterizing a system with slower chemical enrichment.

A useful means of better understanding the nucleosynthesis processes
yielding iron-peak elements 
is their abundance derivation in different
stellar populations.
In the present work, we derive Zn abundances for a sample of 56
bulge field stars, observed at high resolution with the FLAMES-UVES
spectrograph (Lecureur et al. 2007; Zoccali et al. 2006; 
Hill et al. 2011), where we adopt the stellar parameters 
effective temperature T$_{\rm eff}$, gravity log g, 
metallicity [Fe/H]\footnote{We adopted here the usual
spectroscopic notation that [A/B] = log(N$_{\rm A}$/N$_{\rm B}$)$_{\star}$ $-$
log(N$_{\rm A}$/N$_{\rm B}$)$_{\odot}$ and $\epsilon$(A) = log(N$_{\rm
A}$/N$_{\rm B}$) + 12 for the elements A and B.}, and microturbulence 
velocity from these previous determinations.

We compare our results with previous Zn abundance determinations 
for bulge, disk, and halo stars from the literature.
Bensby et al. (2013, and references therein) derived Zn for microlensed
bulge dwarfs. Prochaska et al. (2000), Reddy et al. (2006), 
Nissen \& Schuster (2011),
and Bensby et al. (2014) derived Zn abundances for thick disk stars.
A comparison with thin disk abundances is also given, based on work
by Allende-Prieto et al. (2004), Bensby et al. (2014),
and Pomp\'eia (2003).

We also compare the present results to literature abundances for 
damped Lyman-alpha systems, where Zn is considered as a proxy for Fe.
We included comparisons with analyses by Akerman et al. (2005),
Cooke et al. (2011, 2013), Kulkarni et al. (2007), and Vladilo et al. (2011).

Finally, we also present a model of Zn enrichment in massive spheroid
systems, which is based on the code described in Lanfranchi \& Fria\c ca (2003). It would represent the early bulge enrichment and its chemical evolution.

In Sect. 2 the observations are summarized
and the   adopted atomic constants and solar abundances given. 
In Sect. 3 the basic stellar parameters are listed, and the
abundance derivation of  Zn is described. The results
are compared with literature data and discussed in Sect. 4.
 A summary is given in Sect. 5.


\section{Observations, atomic line parameters, and solar
abundances}

The present UVES data were obtained using the UVES-FLAMES 
instrument at the 8.2 m Kueyen ESO telescope, as described in 
Zoccali et al. (2006), Lecureur et al. (2007), and
Hill et al. (2011).  The spectra cover the wavelength range
4800-6800 {\rm \AA} with a resolution of R $\sim$ 45 000 and
a pixel scale of 0.0147 {\rm \AA}/pix.
Targets are bulge K giants, with magnitudes $\sim$0.5 above the
red clump, in four fields, including Baade's Window.

\begin{figure*}
\centering
\psfig{file=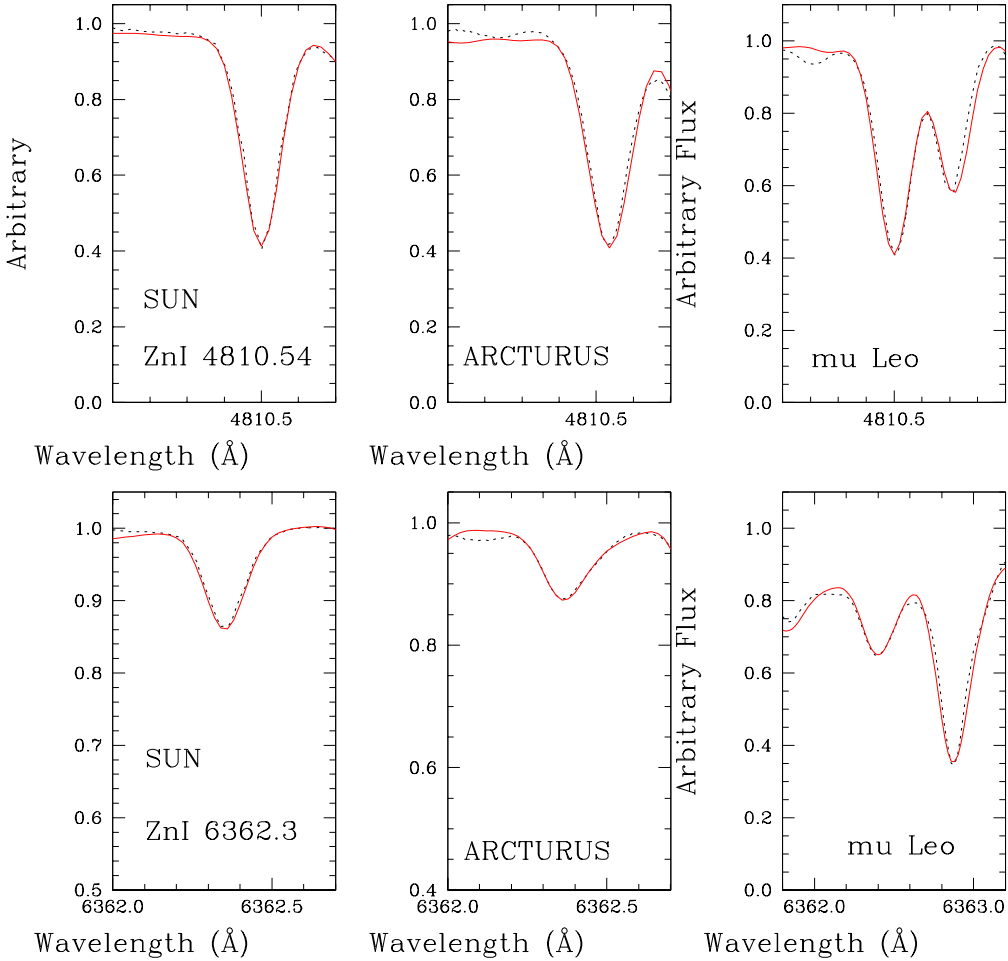,angle=0.,width=14.0 cm}
\caption{Lines of \ion{Zn}{I} fit to the solar
spectrum observed with the UVES spectrograph {and to spectra of}
 Arcturus (Hinkle et al. 2000),
 and $\mu$ Leo observed with the ESPaDOns/CFHT spectrograph. 
  Observed spectrum ({\it dashed lines});
 synthetic
spectra ({\it solid red line}).}
\label{sun} 
\end{figure*}

Zinc lines were checked against oscillator strengths log gf in the literature,
using in particular the data bases from Kur\'ucz (1995) 
website\footnote{http://www.pmp.uni-hannover.de/cgi-bin/ssi/test/kurucz/sekur.html},
NIST\footnote{http://physics.nist.gov/PhysRefData/ASD/lines$_-$form.html}, 
and VALD (Piskunov et al. 1995).
Table \ref{lines2} reports excitation potential, log gf
values from the literature with their references, and adopted
 log gf values.  The lines were fitted to the 
solar high resolution observations using the same UVES spectrograph
\footnote{http://www.eso.org/\-observing/\-dfo/\-quality/\-UVES/\-pipeline/solar{$_{-}$}spectrum.html} as the present sample of spectra, to spectra from Arcturus (Hinkle et al. 2000),
and   from the metal-rich giant $\mu$ Leo, with spectra
 observed with the ESPaDOns/CFHT spectrograph, at a resolution
of R $\sim$ 80 000 and a S/N $\sim$ 500 (Lecureur et al. 2007).
The fits of \ion{Zn}{I} 4810.529 and 6362.339 {\rm \AA} lines
in these reference stars are presented in Fig. \ref{sun}.

\subsection{Autoionization of \ion{Ca}{I} lines}
The measurement of the \ion{Zn}{I} 6362.350 line has to take a continuum lowering in the range
$\sim$6360.8 - 6363.1 {\rm \AA\  into
account},
 owing to the \ion{Ca}{I} 6361.940 auto-ionization line.
Mitchell \& Mohler (1965) identified depressions that are $\sim$2.6 {\rm \AA}
wide at the locations of the \ion{Ca}{I} multiplet lines
at 6318, 6343, and 6363 {\rm \AA}, which are caused by
 auto-ionizing transitions 3d4p 3F$^{\circ}$ - 3d4d 3G.
 Following the recipe that Lecureur et al. (2007) used for the 
6318 {\rm \AA} line, we adjusted the radiative broadening factor for the 6363 {\rm \AA} line to match the line profile to standard stars 
(Sun, Arcturus, $\mu$ Leo), as well as to sample stars, 
thus taking the much reduced lifetime of the level suffering auto-ionization into account. The best-fitting value is 32~000 higher than the standard radiative broadening ($\gamma$$_{\rm rad}$ = 0.21/$\lambda$$^{2}$). Taking this effect into account,
we derived an astrophysical log gf value for the \ion{Ca}{I} 6361.940
line, as reported in Table \ref{lines2}.


\begin{table*}
\begin{flushleft}
\caption{Central wavelengths and oscillator strengths.
References in Column 7: 1 Bi\'emont \& Godefroid 1980; 
2 Ram\'{\i}rez \& Allende-Prieto 2011;
3 Nissen \& Schuster 2011; 4 astrophysical log gf from present fits.}
\label{lines2}      
\centering          
\begin{tabular}{lrrrrrrrrrrrrrrrr}     
\noalign{\smallskip}
\hline\hline    
\noalign{\smallskip}
\noalign{\vskip 0.1cm} 
Species & {\rm $\lambda$} ({\rm \AA}) & \hbox{$\chi_{ex}$ (eV)} & 
\hbox{log gf$_{\rm Kurucz}$} & \hbox{log gf$_{\rm NIST}$} & 
\hbox{log gf$_{\rm VALD2/VALD3}$}  &
 \hbox{log gf$_{\rm literature}$} & \hbox{log gf$_{\rm adopted}$}  \\
\noalign{\vskip 0.1cm}
\noalign{\hrule\vskip 0.1cm}
\noalign{\vskip 0.1cm}
ZnI    & 4810.529 & 4.0782 & $-$0.137 & --- & $-$0.137 &$-$0.17$^1$,$-$0.16$^2$,$-$0.31$^3$ &$-$0.25 \\
    & 6362.339 & 5.7961 & +0.150 & +0.158 & 0.150 &+0.14$^{1,2}$ & +0.05 \\
CaI    & 6361.786 & 4.4510 & +0.954 & --- & +0.317 &$-$0.2$^{4}$ & $-$0.20 \\
CrI    & 4810.509 & 2.9870 & --- & --- & $-$3.142/$-$2.899& $-$2.90$^{4}$   & $-$2.90 \\
TiI    & 4810.705 & 2.4870 & ---  & --- & $-$2.576/$-$2.563& $-$1.00$^{4}$ & $-$1.00 \\
 V1    & 4810.730 & 3.1310 & ---  & --- & $-$1.246/$-$2.534&$-$1.25$^{4}$ & $-$1.25 \\
CrI    & 4810.732 & 3.0790 & ---  & $-$1.30 & $-$1.300/$-$0.644& $-$1.90$^{4}$ & $-$1.90 \\
\noalign{\vskip 0.1cm}
\noalign{\hrule\vskip 0.1cm}
\noalign{\vskip 0.1cm}  
\hline                  
\end{tabular}
\end{flushleft}
\end{table*}

\subsection{Solar abundances}

Table \ref{sol} gives literature abundances for Fe
and Zn for the Sun, Arcturus, and $\mu$ Leo.
The adopted solar abundances for Fe and Zn are from Grevesse \& Sauval
(1998), whereas for Arcturus they are from
 the present fits, when adopting stellar parameters from Mel\'endez et al. (2003). 
For the metal-rich star $\mu$ Leo,
the stellar parameters and C, N, O abundances from Lecureur et al. (2007)
were adopted:
T$_{\rm eff}$ = 4540 K, log g = 2.3, [Fe/H] = +0.3, v$_{\rm t}$ = 1.3
km s$^{-1}$. 
The C, N, O abundances for $\mu$ Leo were revised from
 $\epsilon$(C) = 8.85, $\epsilon$(N) = 8.55, 
$\epsilon$(O) = 9.12, to
[C/Fe]=$-$0.3, [N/Fe]=+0.6, [O/Fe]=$-$0.1 or
  $\epsilon$(C) = 8.55, $\epsilon$(N) = 8.83, 
$\epsilon$(O) = 8.97.
Also for better fitting the line on the red side of
the \ion{Zn}{I} 4810.5 {\rm \AA} line, which is not well fitted 
with literature log gf values,
we refitted them with the astrophysical log gf values reported
 in Table \ref{lines2},
assuming [Cr/Fe]=[Ti/Fe]=0.0 for $\mu$ Leo.
 Finally, for $\mu$ Leo, the resulting Zn abundance
of [Zn/Fe]=$-$0.1 is derived from both ZnI lines.

\begin{table}[h!]
\caption{Solar abundances are from 1: Anders \& Grevesse (1989), 2: Grevesse
 \& Sauval (1998) (adopted), 3: Asplund et al. (2009),
 4: Lodders et al. (2009); Arcturus abundances are from 
(5) Ram\'{\i}rez \& Allende Prieto (2011), and (6) present fits,
computed with parameters from Mel\'endez et al. (2003) (adopted abundances);
$\mu$ Leo abundances are from (7) Lecureur et al. (2007), with Zn abundance
of [Zn/Fe] = $-$0.1, or $\epsilon$(Zn) = 4.80 derived here.
} 
\label{sol}
\[
\begin{array}{lcccccccccccc}
\hline\hline
\noalign{\smallskip}
\hbox{El.} & \hbox{Z} & \multispan6  \hbox{log $\epsilon(X)$} & & & & \\
  \cline{3-11}  \\
\hbox{} & \hbox{} & \multispan4  Sun && \multispan2 
  \hbox{\rm Arcturus}  
&  &  \hbox{\rm $\mu$ Leo} &  \\
  \cline{3-6}  \cline{8-9} \cline{11-11}  \\
&  & (1) & (2)  & (3) & (4) & & (5)  & (6)& & (7) &  \\
\noalign{\smallskip}
\hline
\noalign{\smallskip}
\noalign{\hrule\vskip 0.1cm}
\hbox{Fe} & 26  &~7.67 &  ~7.50  & ~7.50  & ~7.46 & &6.98 
&6.95 & & 7.80 &  \\
\hbox{Zn} & 30  &~4.60 &  ~4.60  & ~4.56 & ~4.65 & & 4.26 & 4.06 & & 4.80 &  \\
\noalign{\vskip 0.1cm}
\hline
\noalign{\smallskip}
\end{array}
\]
\end{table}

\section{Abundance analysis}

Elemental abundances were obtained through line-by-line spectrum synthesis
calculations, carried out 
using the code described in Barbuy et al. (2003) and Coelho et al. (2005).
The molecular lines present in the region, namely the 
CN  B$^2$$\Sigma$-X$^2$$\Sigma$ blue system,
 CN  A$^2$$\Pi$-X$^2$$\Sigma$ red system,
 C$_2$  Swan A$^3$$\Pi$-X$^3$$\Pi$, MgH A$^3$$\Pi$-X$^3$$\Sigma^{+}$,
 and TiO A$^3$$\Phi$-X$^3$$\Delta$ $\gamma$ and
B$^3$$\Pi$-X$^3$$\Delta$ $\gamma$' systems were taken into account.
The atmospheric models were obtained by interpolation in the grid
of MARCS LTE models (Gustafsson et al. 2008). The stellar parameters,
reported in Table \ref{cncorrected}, were adopted from the detailed analyses by
Zoccali et al. (2006) and Lecureur et al. (2007) for 43 bulge field red giants.
Another 13 field red clump stars were analysed by Hill et al. (2011)
based on both the UVES and the GIRAFFE spectra. We adopted the parameters
derived from the UVES spectra, which were not given in Hill et al. (2011) but already
reported in Barbuy et al. (2013).

\subsection{C, N, O abundances and blending CN lines}

The \ion{Zn}{I} 6362.339 {\rm \AA} line is blended with the CN lines
reported in Table \ref{cnlines}, corresponding
to the laboratory measurements by Davis \& Phillips (1963). 
We initially adopted
 the C, N, and O abundances derived in Lecureur et al. (2007).
Based on the observed profile of the \ion{Zn}{I} 6362 {\rm \AA} line, however, 
we realized that in some cases the CN blend was overestimated, producing a
 spurious asymmetry in the red wing of the \ion{Zn}{I} 6362 {\rm \AA} line.
  In these cases (16 stars marked with ** in Table \ref{cncorrected},
 we recomputed C, N, and O abundances.
We finally decided to recompute C, N, and O for all
sample stars, given the influence of the CN blend 
in the Zn abundance derivation from the \ion{Zn}{I} 6362 {\rm \AA}
line.
Oxygen was also derived for the two most metal-poor stars of the sample
BW-f4 and BW-f8, for which no previous derivation
was available. 

\begin{figure*}
\centering
\psfig{file=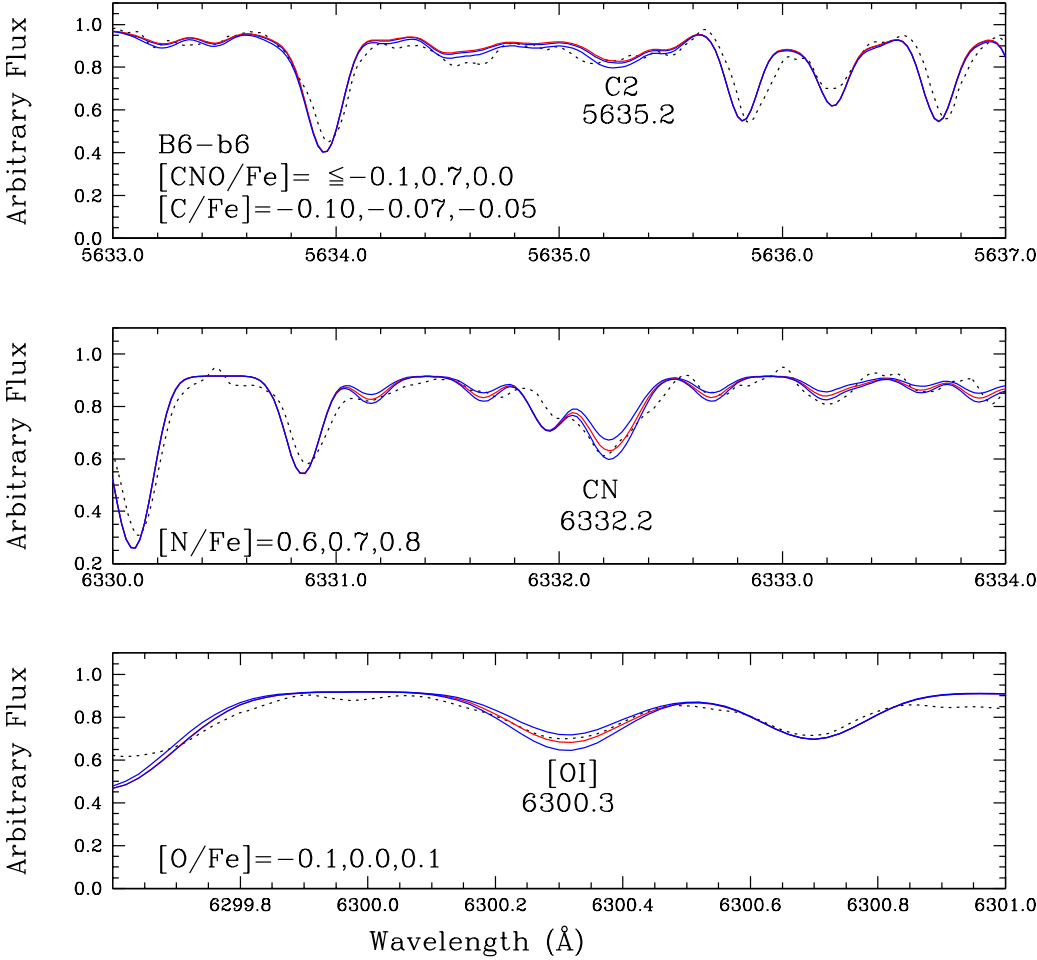,angle=0.,width=14.0 cm}
\caption{CNO in B6-b6: Fits to the Swan C$_2$ (0,1) bandhead at
5635.2 {\rm \AA}, the red CN (5,1) bandhead
at 6332.18 {\rm \AA}, and the forbidden oxygen [OI]6300.31 {\rm \AA}
 line. Dotted line: observed spectrum; red line: synthetic spectra giving
the best fit:
[C/Fe]=$-$0.07, [N/Fe]=+0.7, [O/Fe]=0.0; 
blue lines: synthetic spectra computed with [C/Fe]=$-$0.10 and $-$0.05, 
[N/Fe]= +0.6 and +0.8, and [O/Fe]=$-$0.1 and +0.1.}
\label{cno} 
\end{figure*}

The derivation of C, N, and O abundances was carried out by fitting
the Swan C$_2$ (0,1) A$^3$$\Pi$-X$^3$$\Pi$ bandhead at
5635 {\rm \AA}, the red CN (5,1) 
A$^2$$\Pi$-X$^2$$\Sigma$ bandhead
at 6332.18 {\rm \AA}, and the forbidden oxygen [OI]6300.311 {\rm \AA} line.
Fits to these C,N,O abundance indicators 
 are illustrated in Fig. \ref{cno} for star B6-b6.
These features have to be fitted iteratively, given
that a change in the abundance of any of them has an impact on the
molecule dissociative equilibrium. A further check of the CN
line intensity was done for the asymmetry of the
 \ion{Zn}{I} 6362 {\rm \AA} line.

 In Table \ref{cncorrected} we report the Lecureur et al. (2007)
C, N, O abundances, and in the next column, we assign the letter {\bf c}
when the abundance is confirmed; otherwise, we give the newly
revised abundances.
%
A few comments on individual stars are given in the last column of
Table \ref{cncorrected}:
Telluric means that such lines mask the [OI]6300 and [OI]6363 lines, thus
preventing any possibility of deriving its oxygen abundance, such as in BL-7;
by the comment ``CN-strong'', we  mean
 that CN lines that are too strong blend the \ion{Zn}{I} 6362
{\rm \AA} line, leading to this line being discarded in 15 stars: B6-b2,
B6-f1, B6-f8, BW-b5, BW-f1, BW-f7, B3-b3, B3-b5,
B3-b7, B3-f5, BWc-2, BWc-3, BWc-5, BWc-6, and BWc-8.

No Zn lines were useful for stars B3-b3, B3-f5, and BWc-2,
but they were kept in the line list for reporting their C, N, O abundance
derivation. This means that we derived Zn abundances for 53
(and not 56) stars.
In about a third  of the sample stars,
an asymmetry due to the CN lines was clear, permitting a 
 confirmation of 
the  CNO abundances derived, such as for the star B3-b1.

 These corrected C, N, O abundances were adopted here for
the  purpose of the present paper, which is to
correct the derivation of the Zn abundances. A
 more thorough discussion of these revised oxygen abundances
will be deferred to elsewhere.
The Zn abundances were recomputed for the stars where the CNO abundances
were revised, taking into account the newly derived 
C, N, O abundances, given
in Column 11 of Table \ref{cncorrected}, and the final Zn abundances are
reported in Column 14.

\begin{table}[h!]
\caption{CN lines blending the \ion{Zn}{I} 6362.339 {\rm \AA},
from laboratory measurements by Davis \& Phillips (1963). 
} 
\label{cnlines}
\[
\begin{array}{lccrrrr}
\hline\hline
\noalign{\smallskip}
\hbox{v',v''} & \hbox{$\lambda$({\rm \AA})} & \hbox{Branch}&& 
  \hbox{J}    &   \\
\noalign{\smallskip}
\hline
\noalign{\smallskip}
\noalign{\hrule\vskip 0.1cm}
\hbox{(4,0)} & 6362.743 &R1 & 46  & \\
\hbox{(4,0} & 6362.450  &Q1 & 39  & \\
\hbox{(10,5)} & 6362.765  &P2 & 21 &  \\
\hbox{(10,5)} & 6362.548  &P2 & 10 &  \\
\noalign{\vskip 0.1cm}
\hline
\noalign{\smallskip}
\end{array}
\]
\end{table}

\begin{figure}
\centering
\psfig{file=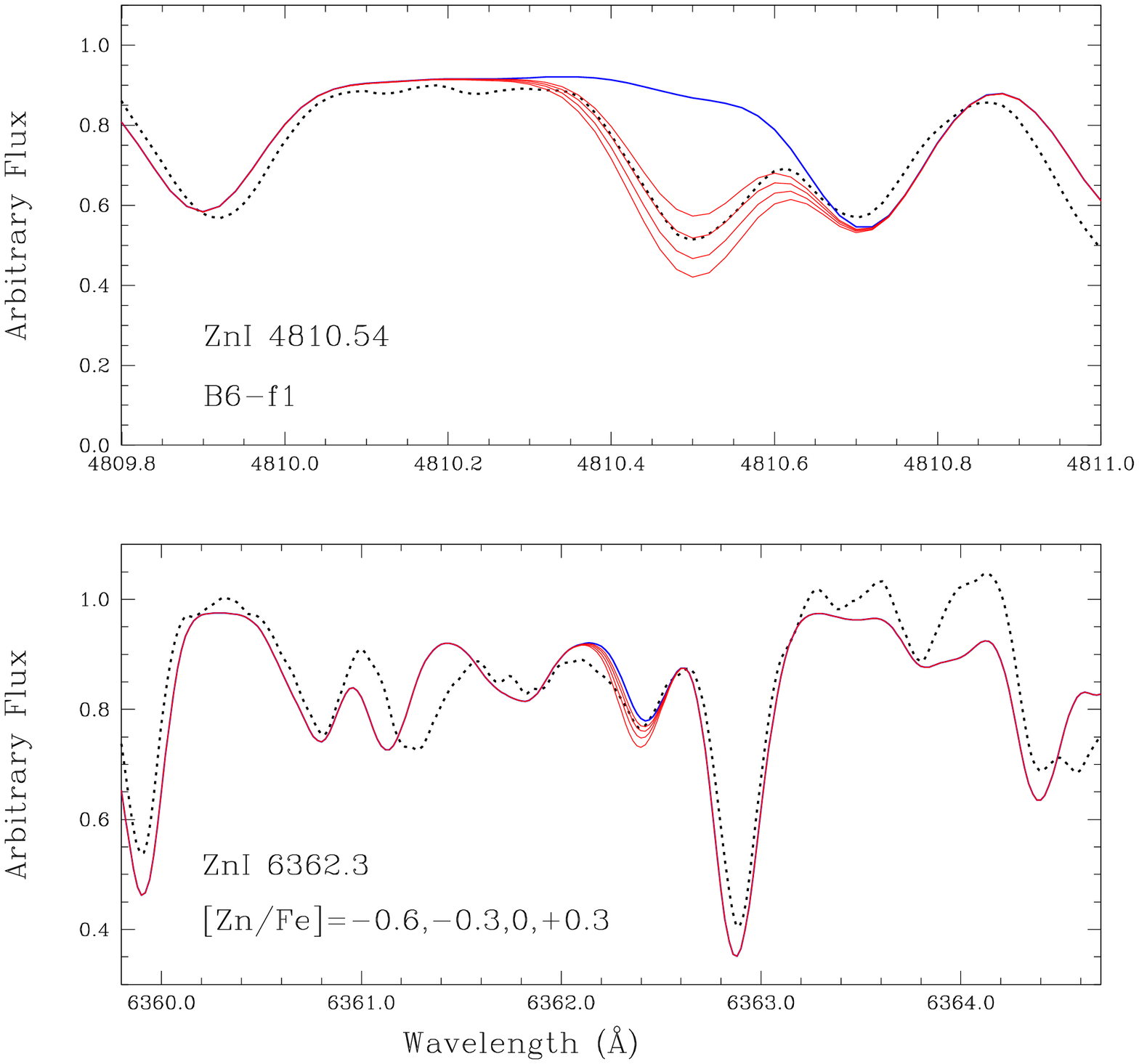,angle=0.,width=9.0 cm}
\caption{\ion{Zn}{I} 4810.54 and 6362.3  {\rm \AA} lines for star B6-f1.
 Dotted lines: observed spectra; red lines: synthetic spectra computed with
the [Zn/Fe] values reported in the lower panels.
The synthetic spectrum in blue has no Zn: for this star the \ion{Zn}{I}
6362.3  {\rm \AA} line was discarded.
}
\label{b6f1zn} 
\end{figure}

\begin{figure}
\centering
\psfig{file=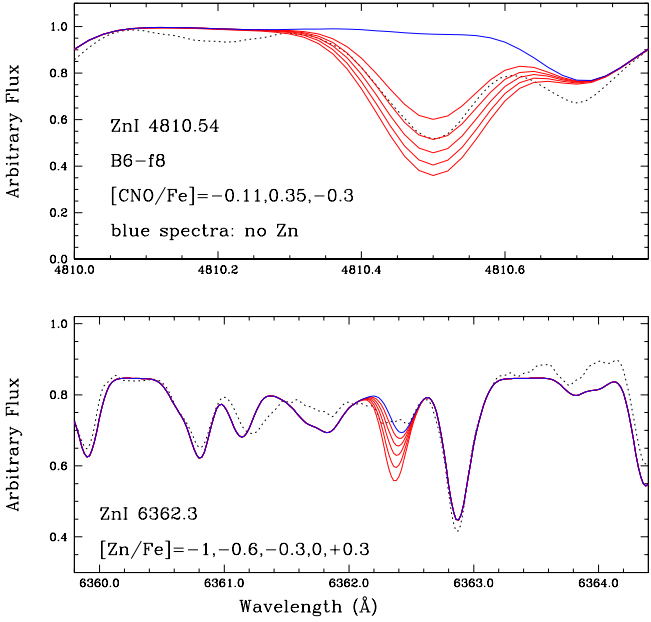,angle=0.,width=9.0 cm}
\caption{\ion{Zn}{I} 4810.54 and 6362.3  {\rm \AA} lines for star B6-f8.
 Dotted red and blue lines: same as in Fig. \ref{b6f1zn}. The \ion{Zn}{I}
6362.3  {\rm \AA} line was discarded
 for this star.}
\label{b6f8zn} 
\end{figure}

\begin{figure}
\centering
\psfig{file=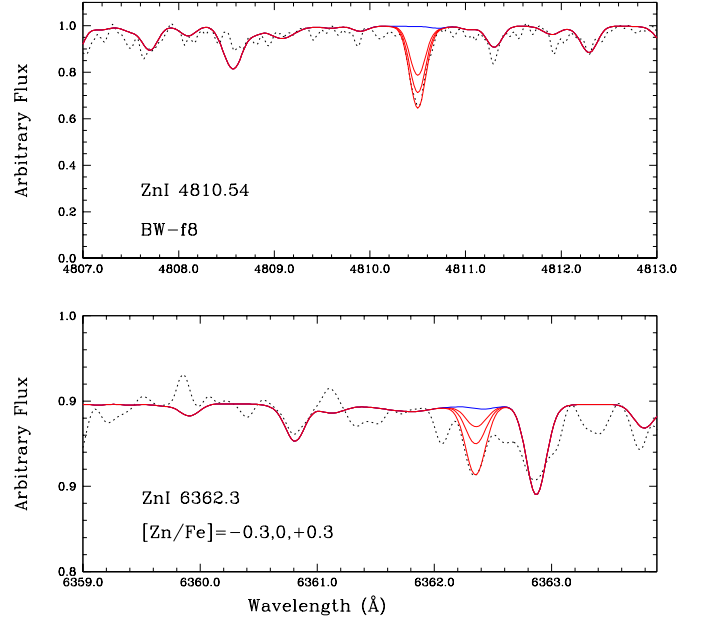,angle=0.,width=9.0 cm}
\caption{\ion{Zn}{I} 4810.54 and 6362.3  {\rm \AA} lines for star BW-f8.
Dotted red and blue lines: same as in Fig. \ref{b6f1zn}.}
\label{bwf8zn} 
\end{figure}

\begin{figure}
\centering
\psfig{file=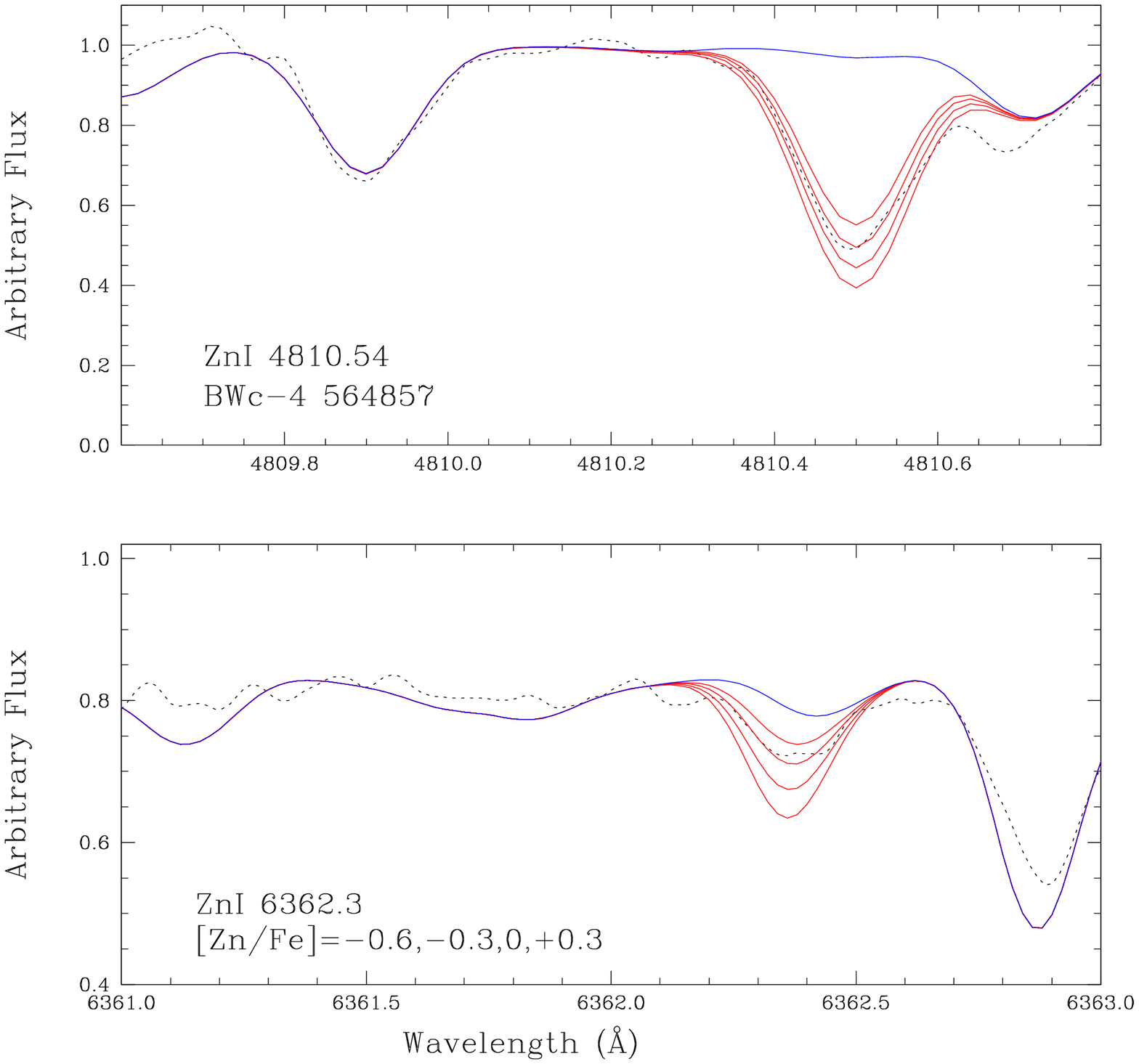,angle=0.,width=9.0 cm}
\caption{\ion{Zn}{I} 4810.54 and 6362.3  {\rm \AA} lines for
red clump star BWc-4.
Dotted red and blue lines: same as in Fig. \ref{b6f1zn}}
\label{bwc4zn} 
\end{figure}

For the fit of the \ion{Zn}{I} 4810 {\rm \AA}
 line, we adopted a procedure of balancing
the continua points at the
pseudo-continuum regions 4808.25, 4811.55, and 4812.6, with more weight
for the 4811.55 {\rm \AA} that appears to be  a better defined continuum.
For the \ion{Zn}{I} 6362.34 {\rm \AA} line, with a few exceptions,
 the local continuum is
affected by the depression due to the \ion{Ca}{I} line and,
despite its being considered in the calculations, some
extra depression remains.
We gave priority to the continuum in the region 6361.5-6362.1 {\rm \AA}
on the blue side of the Zn line.
Examples of fitting ZnI lines in sample spectra are given in Figs.
\ref{b6f1zn}, \ref{b6f8zn}, \ref{bwf8zn}, and \ref{bwc4zn}
for stars B6-f1, B6-f8, BW-f8, and BWc-4. In these figures we also
show the calculations with no Zn (synthetic spectra in blue), 
showing that the \ion{Zn}{I} 4810 {\rm \AA} line is free of blends,
and the  \ion{Zn}{I} 6362 {\rm \AA} line shows a blend with CN lines. 
These figures show examples of 
the cases of no inteference by CN lines (BW-f8), a moderate or negligible
presence of CN lines (BWc-4),
 and strong contaminating CN lines (B6-f1, B6-f8), leading us not
to consider the line for such stars. Star B6-f8 shows a detectable
good fit to the CN line on the red side of the \ion{Zn}{I} 6362 {\rm \AA} line,
but this indicator was discarded as well.

As regards non-LTE corrections,
Takeda et al. (2005) have computed non-LTE effects on both the
\ion{Zn}{I} 4810 and 6362 {\rm \AA} lines used here. 
The corrections are below
0.1 dex for the metal-poor stars and below 0.05 for stars
with [Fe/H] $>$ $-$1.0. Consequently, in the present work, we do not correct
the abundances for this effect.

\subsection{Errors}

 We have adopted 
uncertainties in the atmospheric parameters of $\pm$ 150 K for
effective temperature, $\pm$ 0.20 for surface gravity, and  $\pm$ 0.10 in [Fe/H]
 and $\pm$ 0.10 kms$^{-1}$ for microturbulent velocity, as explained
in Barbuy et al. (2013).

The errors in [Zn/Fe] are computed by using model atmospheres
with parameters changed by these uncertainties, applied to the stars
B6-f1 and BW-f8. The errors are estimated from the differences in [Zn/Fe],
derived using the modified models relative to the adopted model. 
We adopt a mean of the uncertainty on the
 \ion{Zn}{I} 4810 and 6362 {\rm \AA}
 lines.
 These uncertainties are given in Table \ref{errors2} 
for star B6-f1, for which the \ion{Zn}{I} 6362 line is strongly affected 
by the CN blending,
and BW-f8 with negligible CN blending.
The higher sensitivity to effective temperature in B6-f1
is due to the CN lines blending the \ion{Zn}{I} 6362.339 {\rm \AA}
line: it is the CN lines that are more sensitive to temperature. 
Since the stellar parameters are covariant, the sum of these errors
is an upper limit.
On the other hand, a continuum location uncertainty introduces a
further uncertainty in  [Zn/Fe]  of $\pm$0.1.

\begin{table}[ht!]
\caption{Uncertainties on the derived [Zn/Fe] value for model changes of $\Delta$T$_{\rm eff}$ = +150 K,
$\Delta$log g = +0.2, $\Delta$[Fe/H] = +0.1 dex,
 $\Delta$v$_{\rm t}$ = +0.1 km s$^{-1}$,
  and corresponding total error.
} 
\label{errors2}
\[
\begin{array}{lccccc}
\hline\hline
\noalign{\smallskip}
\hbox{Star} & \hbox{$\Delta$T$_{\rm eff}$} & \hbox{$\Delta$$\log$ g}
& \hbox{$\Delta$[Fe/H]} & \hbox{$\Delta$v$_{t}$} & \hbox{($\sum$x$^{2}$)$^{1/2}$} \\
& \hbox{(+150 K)} & \hbox{(+0.2)} & \hbox{(+0.1)} & \hbox{(+0.1 kms$^{-1}$})&\\
\noalign{\smallskip}
\hline
\noalign{\smallskip}
\noalign{\vskip 0.1cm}
\noalign{\hrule\vskip 0.1cm}
\hbox{B6-f1}  &  $-$0.17  &   +0.05 & $-$0.06 & $-$0.01 & 0.19 \\
\hbox{BW-f8}  &     0.00  &   $-$0.05 & +0.10 & $-$0.02 & 0.11 \\
\noalign{\vskip 0.1cm}
\hline
\noalign{\smallskip}
\end{array}
\]
\end{table}

\section{Results}

\begin{figure*}
\centering
\psfig{file=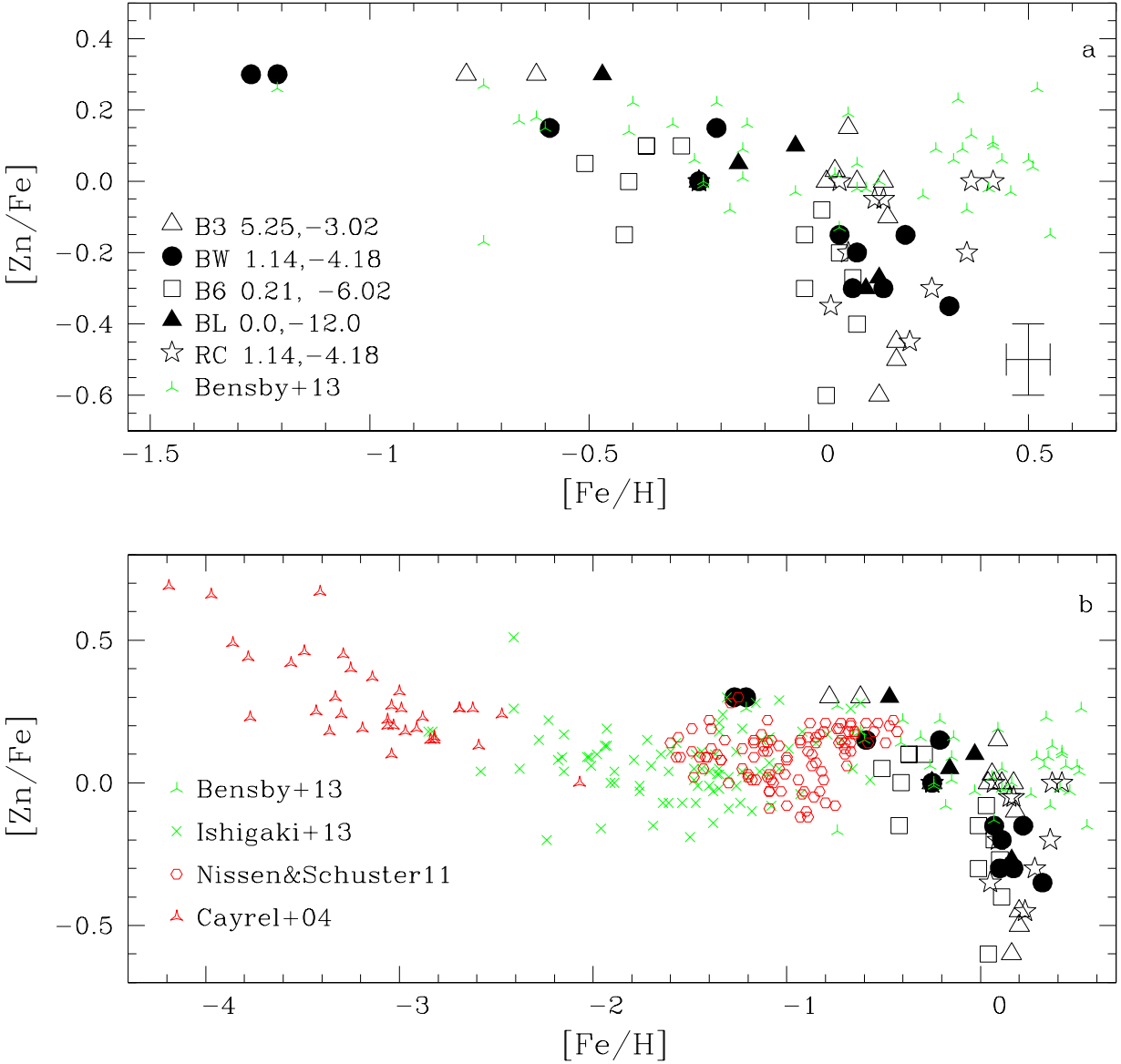,angle=0.,width=17.0 cm}
\caption{[Zn/Fe] vs. [Fe/H] for present results compared with
literature abundances for halo and bulge stars. 
Panel (a): plotted in the metallicity range $-$1.5$<$[Fe/H]$<$+0.7,
corresponding to that of sample bulge giant stars, compared with 
the bulge dwarfs by Bensby et al. (2013). The subsamples are indicated
by different symbols and identified by the central field (l,b) values;
Panel (b): plotted for
$-$4.4$<$[Fe/H]$<$+0.7, encompassing halo stars from Cayrel et al. (2004),
halo and thick disk stars from Ishigaki et al. (2013) and Nissen \& 
Schuster (2011), and again the bulge dwarfs by Bensby et al. (2013).
The symbols indicate:
 Present data in black: open triangles: NGC 6553 field (designated B3),
filled circles: Baade's Window field (BW); open square: field at $-$6$^{\circ}$ 
(designated B6); filled triangles: Blanco field (designated BL);
open stars: Red clump stars in Baade's Window (designated RC). Literature
data: red three-pointed stars: Cayrel et al. (2004); 
green crosses: Ishigaki et al.
(2013); red open circles: Nissen \& Schuster (2011);
 green 3-pointed stars: Bensby et al. (2013). A representative error bar
is given in the lower right corner of the upper panel.}
\label{plotzn} 
\end{figure*}

The derived Zn abundances are given in Table \ref{cncorrected}.
In Fig. \ref{plotzn}, the [Zn/Fe] vs. [Fe/H] behaviour is shown
for the present sample, where different symbols represent the four different
fields, and the red clump stars.
In Fig. \ref{plotzn}a, we overplotted 
 the Zn abundances of dwarf bulge stars observed thanks
to microlensing magnification, by Bensby et al. (2013).
There is good agreement between the present
results and those by Bensby et al. at metallicities $-$1.4 $<$ [Fe/H] $<$ 0.0. 
 On the other hand, for our metal-rich giants with [Fe/H] $>$ 0.0, we find
a wide spread of $-$0.6 $<$ [Zn/Fe] $<$ +0.15 and a decreasing trend
with metallicity. Further comments on the Bensby et al. sample
of dwarf bulge stars are given in Sect. 4.3.
In Fig. \ref{plotzn}b we show the same stars as in Fig. \ref{plotzn}a,
adding metal-poor halo stars analysed by Cayrel et al. (2004)
and halo and thick disk stars results from
Ishigaki et al. (2013) and Nissen \& Schuster (2011).
Ishigaki et al. results seem to show lower [Zn/Fe] in the
-3.0 $<$ [Fe/H $<$ $-$1.0 metallicity range with respect
to the Nissen \& Schuster and Cayrel et al. results.

\subsection{Comparison with thick disk Zn abundances}
\begin{figure*}
\centering
\psfig{file=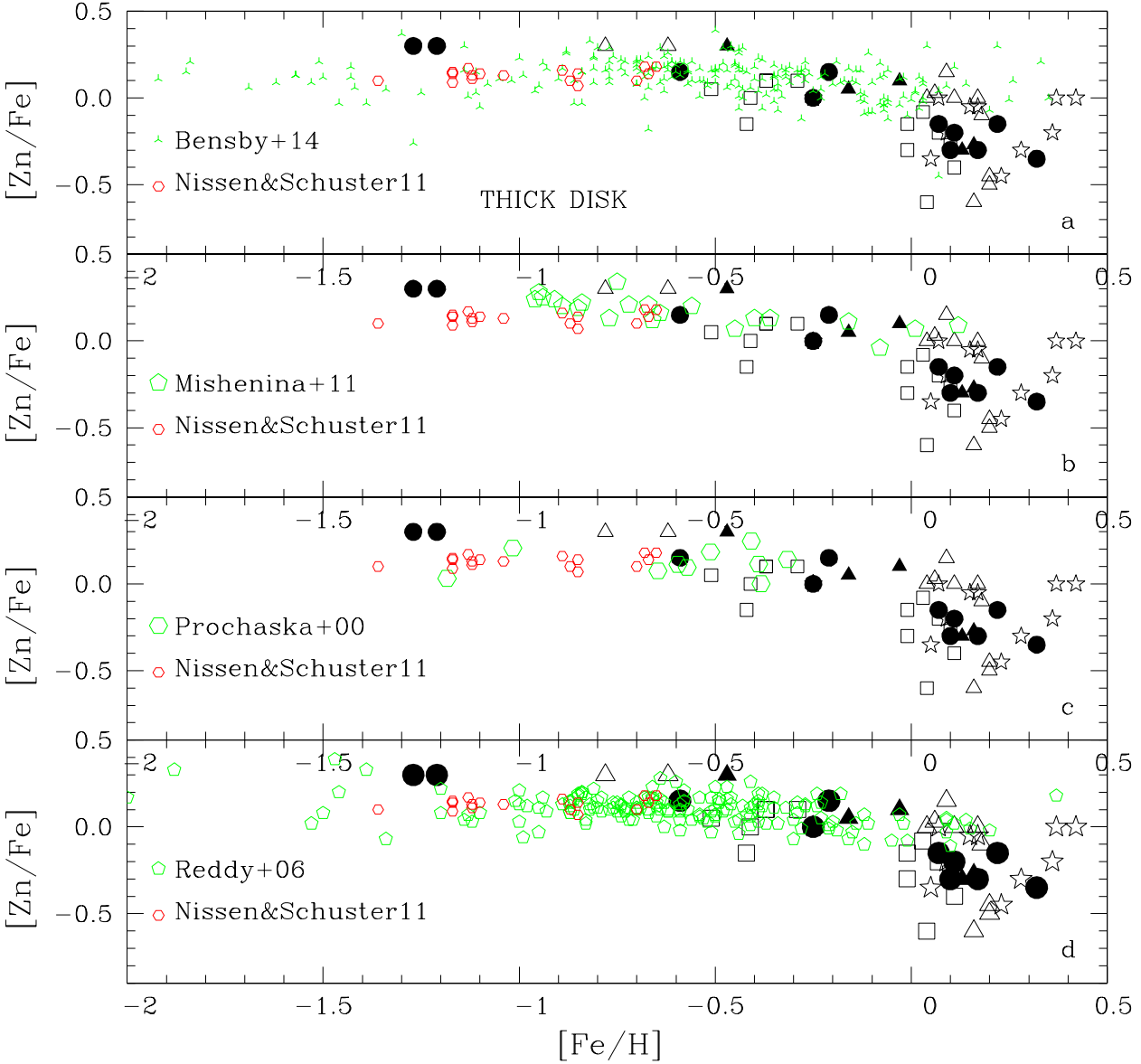,angle=0.,width=17.0 cm}
\caption{[Zn/Fe] vs. [Fe/H] for present results
(same symbols as in Fig. \ref{plotzn}), compared with
the thick disk  
analyzed by Nissen \& Schuster (2011)  (maintained in all panels in this
figure), and  the thick disk data by a: Bensby et al. (2014); 
b: Mishenina et al. (2011);
c: Prochaska et al. (2000); d Reddy et al. (2006).}
\label{plotzn1} 
\end{figure*}

In Fig. \ref{plotzn1} we compare the present bulge Zn abundances
with those for thick disk stars in the literature.
 The results by Nissen \& Schuster (2011)
are maintained in all panels in Fig.  \ref{plotzn1},
showing good agreement with other samples of thick disk stars.
Given the narrow spread and the metallicity range covered by these data,
we adopted the Nissen \& Schuster (2011) results
 as a reference for the thick disk.
Panels a, b, c, d show the Zn abundances derived respectively by
 Bensby et al. (2014), Mishenina et al. (2011), 
Prochaska et al. (2000), and Reddy et al. (2006).
 All thick disk samples show
very good agreement with the present data, in the
metallicity range of thick disk stars (-1.2 $\simless$ [Fe/H] $\simless$ 0.0).
The Bensby et al. (2014)
thick disk stars were selected based on kinematical criteria
TD/D $>$ 2, with these numbers corresponding to a probability
  defined in Bensby et al. (2003).
Thick disk stars from Bensby et al. (2014) displayed in Fig. 
\ref{plotzn1}a include  an old metal-poor  thick disk,
 whereas the  more metal-rich thick disk stars are younger ($<$8Gyr).
The Bensby et al. old metal-poor thick disk stars,
reaching down to [Fe/H]$\approx$-2.0, show a moderate enhancement
of [Zn/Fe]$\approx$+0.15, compatible with our results for 
the metal-poor bulge stars.  Their more metal-rich thick disk component
with [Fe/H] $>$ $-$0.3
show Zn-to-iron essentially solar, whereas our sample includes stars
with low Zn-to-iron.
The metallicity limits for the thick disk often give
[Fe/H]=$-$0.3 as an upper limit; however, Bensby et al. classify some of
their more metal-rich stars as probable thick-disk members, based on
kinematics, as explained above, and they tend to be younger than 
$\simless$ 8 Gyr.
 Comments on the Bensby et al. sample consisting of dwarfs
are given below in Sect. 4.3.

In the metallicity range
  occupied by the old thick disk stars,
 the bulge appears compatible with the thick disk abundances
for the samples from Bensby et al. (2014),
Mishenina et al. (2011), Prochaska et al. (2000), and Reddy et al.
(2006). A distinction between the present data and thick disk
stars may be present for the metal-rich stars,
 where most stars in the present sample
show Zn under-abundances, and the Bensby et al. data show
[Zn/Fe]$\sim$0.0. It is important to note that 
there are not as many such 
old metal-rich stars in the thick disk population
studied by Bensby et al. (2013).

\subsection{Comparison with thin disk Zn abundances}

\begin{figure*}
\centering
\psfig{file=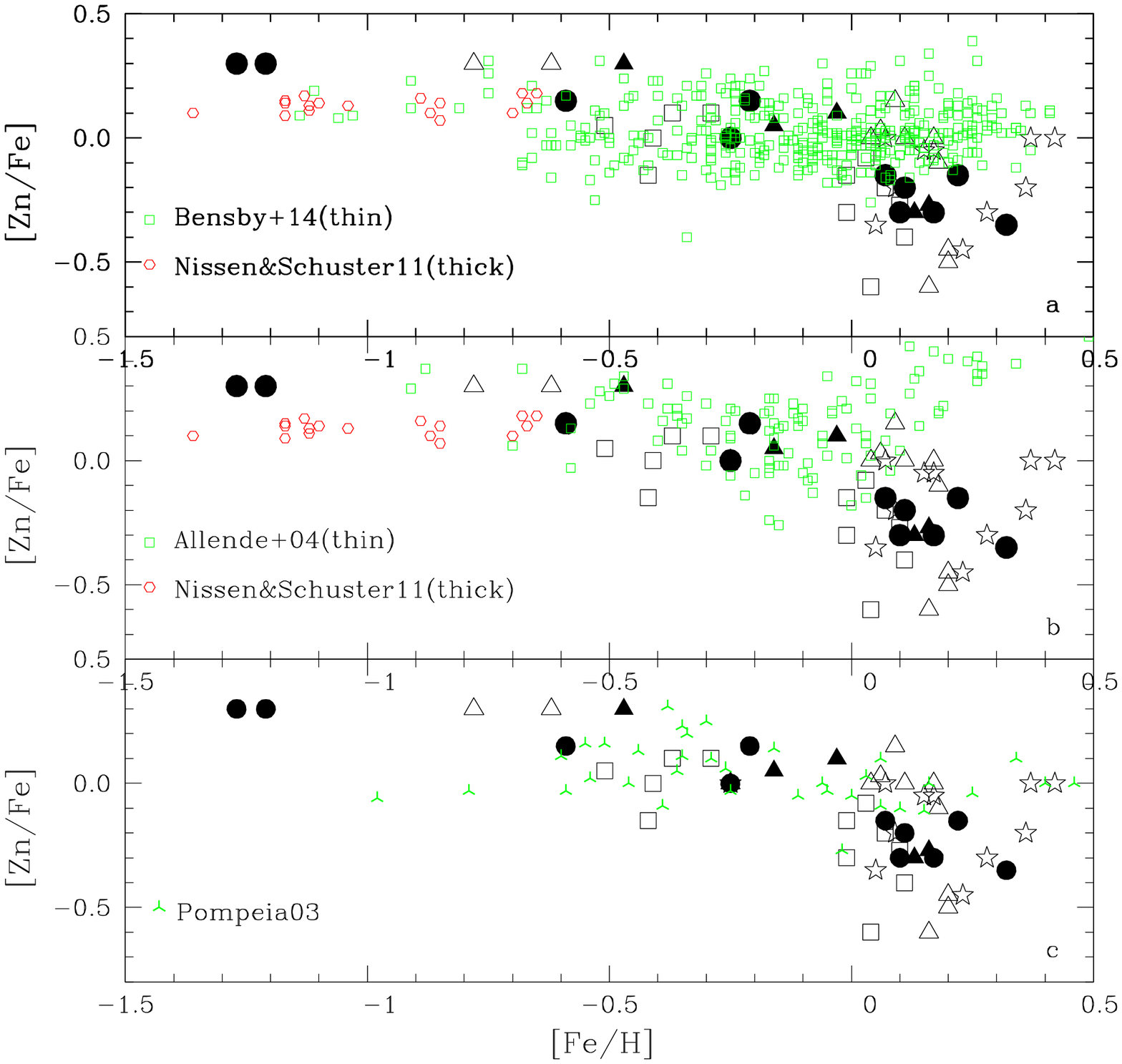,angle=0.,width=17.0 cm}
\caption{[Zn/Fe] vs. [Fe/H] for present results compared with
thin disk stars by a: Bensby et al. (2014);
b: Allende-Prieto et al. (2004); c: Pomp\'eia (2003).
Thick disk stars by  Nissen \& Schuster (2011) are also shown in panels a and b.}
\label{plotzn3} 
\end{figure*}

In Fig. \ref{plotzn3} we compare the present Zn abundances
with those for thin disk stars derived by Bensby et al. (2014),
and Allende-Prieto et al. (2004).
The data are also compared with the metal-rich dwarf stars
from  Pomp\'eia (2003).
 Thick-disk stars by Nissen \& Schuster (2011)
are also kept in all plots, as a reference for the thick disk.
 The thin disk Zn abundances
from Bensby et al. (2014) are consistent with a mean [Zn/Fe]$\sim$0.0, with
 a fraction of stars showing higher [Zn/Fe] up to $\sim$+0.4. 
The Allende-Prieto et al. (2004) stars show a mean [Zn/Fe]$>$0 and 
a clear and an increasing trend with metallicity.
The dwarf stars studied by Pomp\'eia (2003) overlap with the present
bulge stars data, except for the more metal-rich ones,
which show no under-abundance. This difference may indicate that
the Pomp\'eia (2003) sample would be rather old thin disk stars 
and not bulge-like stars, as concluded
in Trevisan et al. (2011) and
again discussed in Trevisan \& Barbuy (2014).

\subsection{Inspecting differences with literature samples}

The comparisons with literature data in the previous sections
include several samples of dwarf stars:
the  microlensed dwarf bulge stars from Bensby et al. (2013),
the thin and thick disk stars from Bensby et al. (2014),
the thin disk stars from Allende-Prieto et al. (2004)
and Pomp\'eia (2003), and the halo and thick disk stars from
Nissen \& Schuster (2011).
Bensby et al. (2013, 2014) and Allende-Prieto et al. (2004)
 used the \ion{Zn}{I} 4810.5
and 6362.2 {\rm \AA} lines, and
Nissen \& Schuster (2011) used the \ion{Zn}{I}  4722.1 and 
4810.5 {\rm \AA} lines.
(Pomp\'eia (2003) does not report the lines used.)

The lines used are, in most cases, shared with the present work.
The dwarfs are all hotter than the present sample, such
that their samples have spectra with essentially no molecular lines.
This could explain any difference in abundances from the \ion{Zn}{I}
 6362.2 {\rm \AA} line, that has a blend with CN lines affecting
our determinations. On the other hand, the \ion{Zn}{I} 4810.5 {\rm \AA}
line has no molecular lines, and it shows the same Zn
deficiencies as the \ion{Zn}{I} 6362.2 {\rm \AA} line for many of
the most metal-rich giant stars, but differently from the dwarfs.

The differences for metal-rich stars could stem from other
blends, possibly unknown. At present we cannot identify a reason
not to confirm the
Zn-under-abundance in some of the present sample of metal-rich bulge giants. 
The flat or decreasing [Zn/Fe] at high metallicities has the important
consequence of indicating (or not) the contribution of SNIa (see Sect. 4.5).



\subsection{Comparison with damped Lyman-$\alpha$ systems}

 A comparison of the present bulge data with DLAs is possible thanks
to the availability of a large data base of Zn abundances over a
 wide range of metallicities for these objects. Such
 a comparison can shed light not only on the nature of DLAs, 
but also on the formation process of bulges.

The high H I column densities that characterize the DLAs indicate that they belong to environments with relatively high gas density, which could give rise to components of present day massive galaxies. The presence of metals in their spectra is also a clue that star formation has already taken place at significant rates inside them or in their neighbourhood.  In this way,  two favoured candidates for sites hosting DLAs are proto-disks and young star-forming spheroids.
Lanfranchi  and Fria\c ca (2003) have investigated the evolution of the metallicity of DLAs in order to set constraints on the nature of these objects and also to shed light on the connections  between DLAs and galaxy formation. The comparison of the observed trends of [$\alpha$/Fe] and [N/$\alpha$] with the predictions of a chemo-dynamical model is consistent with a scenario in which the DLA population is dominated by disks at low redshifts and by young spheroids at high redshifts. 
With the data base available for that work,
nearly the totality of DLAs at $z > 1.5$ are explained as spheroid systems
formed in the redshift range 1.7$<$z$<$4.5,
with  masses between 10$^9$ and 10$^{10}$ M$_{\odot}$
and typical  specific star formation 
rates of 1 to 3 Gyr$^{-1}$,
 where the definition of specific star formation
is given in the footnote below\footnote{Specific star formation rate (SFR)
  (in  Gyr$^{-1}$) is the ratio of the SFR
 in M$_{\odot}$ Gyr$^{-1}$  over the gas mass in M$_{\odot}$ 
 available  for star formation.}.
These objects might have taken part in merger processes that then lead to bulges. It is possible that the most massive of them, those
with N$_{\rm HI} \approx 10^{22}$ cm$^{-2}$, could be proto-bulges.


Zinc has played a special role in estimations of the metallicity of DLAs 
(Pettini et al. 1990, 1994, 1997; Akerman et al. 2005). 
That Zn is non-refractory
 guarantees that it is not heavily depleted in the ISM. 
On the observational side,
 Zn has two strong transitions at 2026 and 2062 {\rm \AA},
which almost always lie outside the Ly-$\alpha$ forest
 and are rarely saturated owing to their low oscillator strengths
 and low Zn abundances. 
Given that Zn is only a trace element that contributes $\approx 10^{-4}$ of
the mass density for the heavy elements, and because of the weakness of the Zn
transitions, it is impossible to measure its abundance
 in low-metallicity DLAs, and besides that,
 the large rest-frame wavelengths of the transitions
 at 2026 and 2062 {\rm \AA} make them difficult to
measure at high redshift. 

Studies of Zn in damped Lyman-$\alpha$ systems with different redshifts
have shown that [Zn/H] tends to show no correlation with
distance and age. Akerman et al. (2005), for example, suggest
that [Zn/H] in DLAs shows a constant value of [Zn/H] = $-$0.88$\pm$0.21
in the redshift range 1.86 $<$ z$_{abs}$ $<$ 3.45. Comparisons of stellar data with abundances in DLAs were carried out
by several authors (Pettini et al. 1994; Pettini et al 1997a;
Prochaska et al. 2000; Wolfe et al. 2005; Nissen et al. 2007).

\begin{figure*}
\centering
\psfig{file=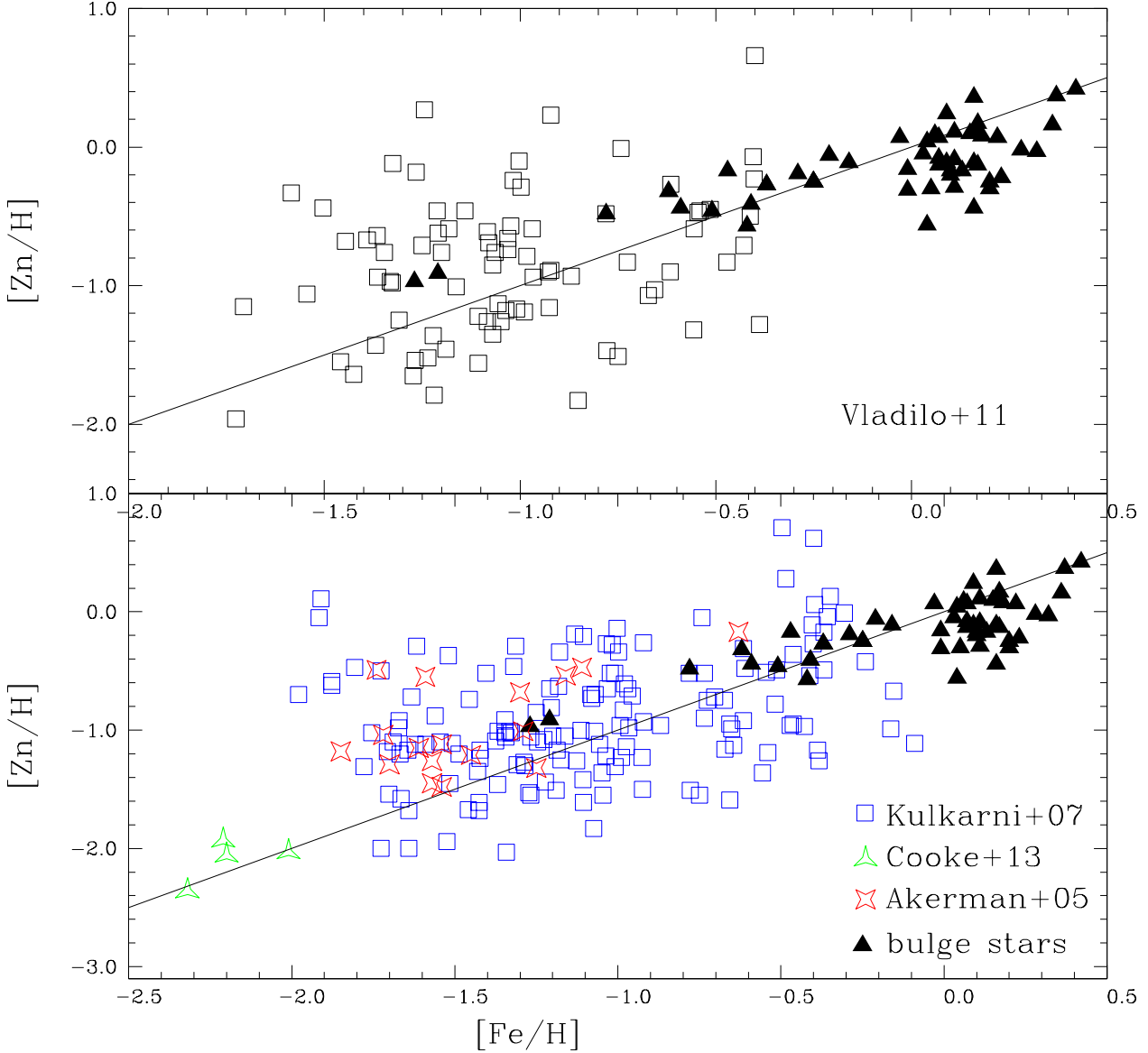,angle=0.,width=17.0 cm}
\caption{[Zn/H] vs. [Fe/H] for the present sample (black filled triangles)
and for the DLA samples by: (upper panel) Vladilo et al. (2011;
 open squares);
(lower panel)  Kulkarni et al. (2007; open squares), 
using a tranformation [Fe/H]
vs. redshift from Pei \& Fall (1995), 
and Cooke et al. (2013; green 3-pointed stars), Akerman et al. 
(2005; red four-pointed stars). A X=Y line is drawn.
}
\label{DLA} 
\end{figure*}

In Figs. \ref{DLA}a,b, we plot [Zn/H] vs. [Fe/H] for the present sample
and for the DLA samples by  Akerman et al. (2005), Cooke et al. (2011, 2013), 
 Kulkarni et al. (2007), and  Vladilo et al. (2011).
For the Akerman et al. (2005) data, we did not consider the cases with
upper limits alone.
 For the DLA samples of  Kulkarni et al. (2007) and
  Vladilo et al. (2011),
 a tranformation [Fe/H] vs. redshift from the literature
 (Pei \& Fall 1995) was adopted.
Other such relations are given by Cen \& Ostriker (1999)
and Madau \& Pozzetti (2000), for example. 
This figure shows that a comparison of
Zn in DLAs and in bulge stars benefit from an overlap only at metallicities
$-$1.5 $<$ [Fe/H] $<$ $-$0.1, with most sample bulge stars being more metal-rich
than this.

\subsubsection{Selected DLAs with Fe abundance measurements}

\begin{table}[h!]
\caption{DLA systems for which Fe abundances were measured,
to which dust depletion corrections were applied. 
Columns [Fe/H]$_{\rm c}$ and [Zn/Fe]$_{\rm c}$ correspond to
Fe abundance dust-depletion-corrected values.
} 
\label{DLA-fe}
\[
\begin{array}{lccr@{}r@{}r@{}r@{}r@{}}
\hline\hline
\noalign{\smallskip}
\hbox{QSO} & \hbox{z$_{\rm abs}$} & \hbox{[Zn/H]}
& \hbox{[Fe/H]} & \phantom{-}\hbox{[Fe/H]$_{\rm c}$} & 
\phantom{-}\hbox{[Zn/Fe]} 
& \phantom{-}\hbox{[Zn/Fe]$_{\rm c}$}    &   \\
\noalign{\smallskip}
\hline
\noalign{\smallskip}
\noalign{\hrule\vskip 0.1cm}
\noalign{\smallskip}
\noalign{Akerman et al. 2005}
\noalign{\smallskip}    
\noalign{\hrule\vskip 0.1cm}
\noalign{\smallskip}
\hbox{B0438-436} & 2.34736 & $-$0.68 & $-$1.30& $-$0.87& 0.62 &  0.24&\\        
\hbox{B0458-020} & 2.03950 & $-$1.15 & $-$1.61& $-$1.34& 0.46 &  0.21&\\        
\hbox{B0528-250} & 2.14100 & $-$1.45 & $-$1.57& $-$1.34& 0.12 & $-$0.09&\\      
\hbox{B0528-250} & 2.81100 & $-$0.47 & $-$1.11& $-$0.57& 0.64 &  0.18&\\        
\hbox{B1055-301} & 1.90350 & $-$1.26 & $-$1.57& $-$1.32& 0.31 &  0.08&\\        
\hbox{B1230-101} & 1.93136 & $-$0.17 & $-$0.63&  0.10& 0.46 & $-$0.15&\\        
\hbox{B2314-409} & 1.85730 & $-$1.01 & $-$1.29& $-$1.01& 0.28 &  0.02&\\        
\noalign{\smallskip}
\noalign{\hrule\vskip 0.1cm}
\noalign{\smallskip}
\noalign{Vladilo et al. 2011}
\noalign{\smallskip}
\noalign{\hrule\vskip 0.1cm}
\noalign{\smallskip}
\hbox{0216+080}  & 2.2930 & $-$0.63 &$-$1.12&$-$0.67 &0.49 &0.10 &\\
\hbox{2206-199A} & 1.9200 & $-$0.33 &$-$0.87&$-$0.25 &0.54 &0.02 &\\
\noalign{\vskip 0.1cm}
\hline
\noalign{\smallskip}
\end{array}
\]
\end{table}

For a few DLAs studied by Akerman et al. (2005) and Cooke et al. (2013),
iron abundances ([Fe/H]) were measured, in addition to zinc abundances.
We selected these objects to analyse the relation between the abundances
 of iron and zinc in DLAs.
Vladilo et al. (2011) quote only [Zn/H] values for most of their DLA sample.
However, they were able to derive, based on archive UVES spectra, 
the \ion{Fe}{II} column densities for two DLAs of their \ion{Zn}{II} + \ion{S}{II} sample ,
0216+080  at $z_{abs}$=2.2930, and 2206-199A  at $z_{abs}$=1.9200,
which we also include in our list of measured Fe abundances in DLAs. 
An important concern is that Fe is a refractory element, 
and its depletion from the
gas phase (observed through the absorption lines) into dust
has to be taken into account.
Therefore we applied a dust correction to the [Fe/H] values  of
the Akerman et al. (2005) and Vladilo et al. (2011) subsamples.
No dust correction was needed for the Cooke et al. data,
because their systems have [Fe/H] $<$ $-$2, and at these low metallicities
dust depletion becomes negligible (Pettini et al. 1997a).
The systems for which we apply dust-depletion corrections for Fe are
listed in Table \ref{DLA-fe}.

Applying dust corrections to the chemical abundances of a gas system,
either a DLA or the Galactic ISM,  is a very complex task. 
We considered a variety of dust correction models, 
following Lanfranchi \& Fria\c ca (2003),
to obtain the depletion $\delta_X$ of a given element X in the DLA,
thus recovering the intrinsic abundance [X/H]$_i$ of that element
from the observed one [X/H]:
\begin{equation}
\delta_X \;=\; [X/H] \; -\; [X/H]_i
.\end{equation}

In the Galactic ISM, 
the suprasolar [Zn/Fe], [Zn/Cr], and [Zn/Si] ratios provide evidence
of dust depletion by the refractories Fe, Cr, and Si.
In addition, the dust depletion pattern depends on the type of environment
through which the line of sight passes.
Savage \& Sembach (1996) considered four different dust depletion patterns 
corresponding to four types of Galactic ISM environments:
(1) cool clouds in the Galactic disk (CD),
(2) warm clouds in the disk (WD), (3) disk plus halo clouds (WHD), 
and (4) warm halo clouds (WH).
On the other hand, we should consider 
a nucleo-synthetic contribution to the Zn/Fe enhancement.

In our dust corrections to the Fe abundances, we followed 
Lanfranchi \& Fria\c ca (2003).
We applied four distinct dust models to 16 DLAs
of Lanfranchi \& Fria\c ca (2003) with small uncertainties for the abundances of
zinc, iron, and one more refractory element (Cr, Si, or Mg).
We assumed a range for the intrinsic [Zn/Fe] ratio of 0.0 (solar),
 0.1, 0.2, and 0.3.
The suprasolar values of [Zn/Fe] are suggested by determinations of low metallicity objects
and could be checked {\it \emph{a posteriori}}.
We  take the WD and the WH as reference media. 
It is not appropriate to consider the cool clouds as a reference for
the DLA environment, because the cool clouds exhibit levels of dust depletion
([Fe/Zn], [Cr/Zn], [Si/Zn]) that are much higher than those observed in DLAs,
and the molecular hydrogen fractions are typically low in DLAs, 
in contrast to the large number of molecules in the cool clouds.
In addition, by using models of chemical evolution,
Vladilo et al. (2011) have shown that, for most of the DLAs in their sample,
the iron depletion is within the range delimited by the values typical 
of WD and WH Galactic clouds.

\begin{figure*}
\centering
\psfig{file=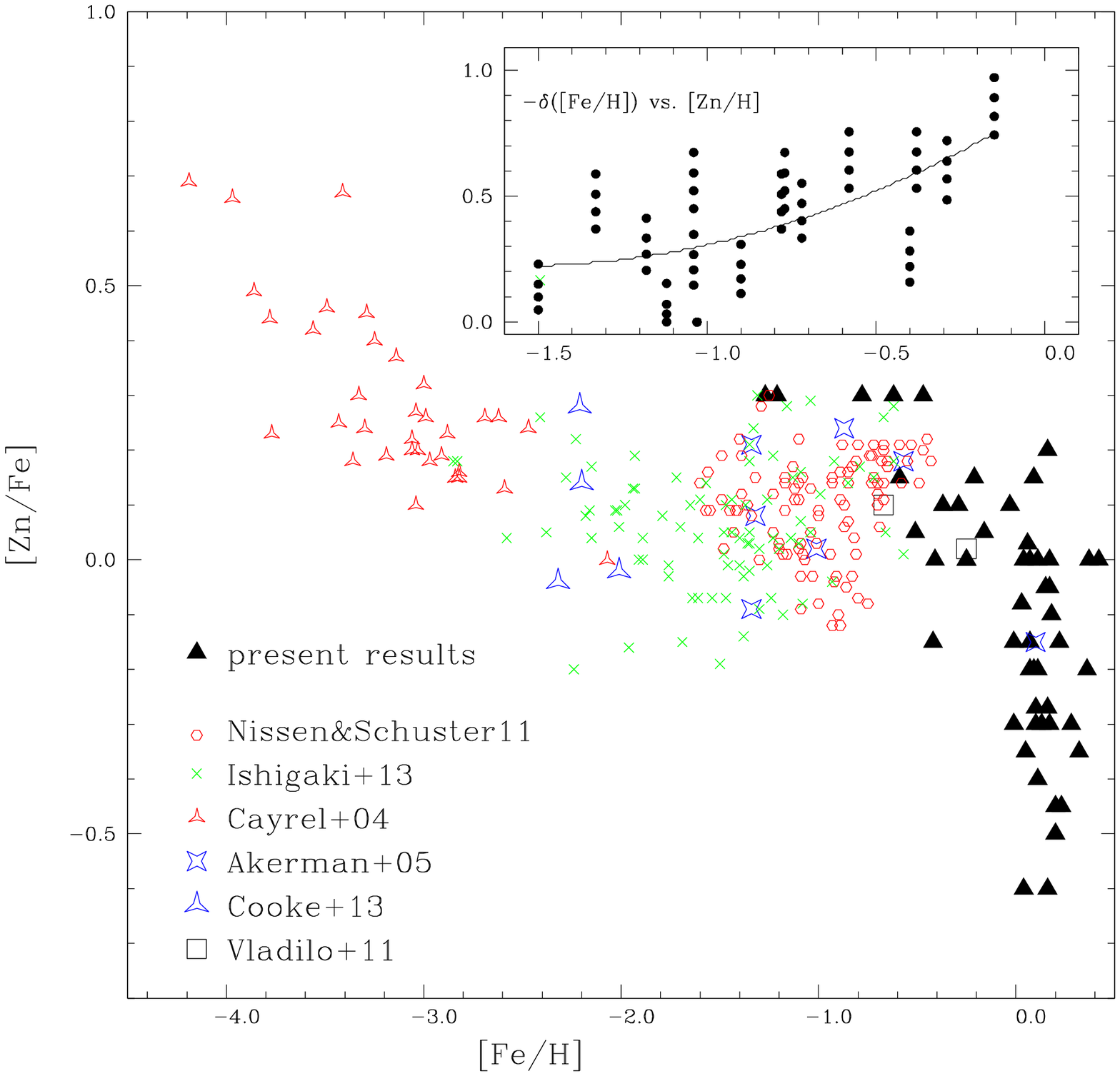,angle=0.,width=17.0 cm}
\caption{[Zn/Fe] vs. [Fe/H] for the present sample
and for the DLA samples by  Akerman et al. (2005) and
Cooke et al. (2013). We also plot the halo stars from
Cayrel et al. (2004) and thick disk and halo stars data
from Nissen \& Schuster (2011) and Ishigaki et al. (2013).
 The inserted plot shows the trend for the dust depletion correction
 for the abundance of iron with the observed [Zn/H] derived 
from applying four distinct dust models to 16 DLAs.
The curve is the quadratic fit to the trend that is used for dust correction.
The abundances from the model calculations are averaged values 
inside several shells,
shown as curves labelled according to the inner and the outer radius
 of each shell.
}
\label{DLAfe} 
\end{figure*}

The degree of depletion should increase with metallicity.
We cannot use [Fe/H] directly to obtain the metallicity
because iron itself is highly depleted into dust.
Therefore, [Zn/H] is taken as a metallicity indicator because it is a volatile element.
We  derive a correction for dust depletion as a function of metallicity
from the trend of the iron depletion predicted by the dust models
with respect to the observed [Zn/H] values,
applying the four distinct dust models to 16 DLAs.
Then, a quadratic fit to the resulting total of 64 points gives
the dust correction as a function of [Zn/H].
 The inserted plot Fig. \ref{DLAfe} shows the trend of the dust depletion 
correction for the abundance of iron
with the observed [Zn/H] derived from applying our dust
 models to the DLAs selected from
Lanfranchi \& Fria\c ca (2003).
The curve is the quadratic fit to the trend that is used for dust correction.

Finally, although less sensitive to dust depletion, the zinc abundance
is also corrected in the way described in Lanfranchi \& Fria\c ca (2003).
Figure \ref{DLAfe} shows [Zn/Fe] vs. [Fe/H] for
the DLA samples with determination of both [Zn/H] and [Fe/H],
compared with the present results for bulge stars,
thick disk, and halo stars from Nissen \& Schuster (2011)
 and Ishigaki et al. (2011) and
for metal-poor halo stars from Cayrel et al. (2004).
The Akerman et al. (2005) and Vladilo et al. (2011) data have been corrected
for dust depletion as explained above.

The Zn vs. Fe enrichment indicated by the DLA data from
 Akerman et al. (2005) and Vladilo et al. (2011) 
 shows an overlap not only with
thick disk and halo stars data from Nissen \& Schuster (2011)
 and Ishigaki et al. (2013),
but also with the pattern of the bulge stars,
including  a few subsolar [Zn/Fe] values.

\subsection{Chemical evolution models of zinc in massive spheroids}

\begin{figure*}
\centering
\psfig{file=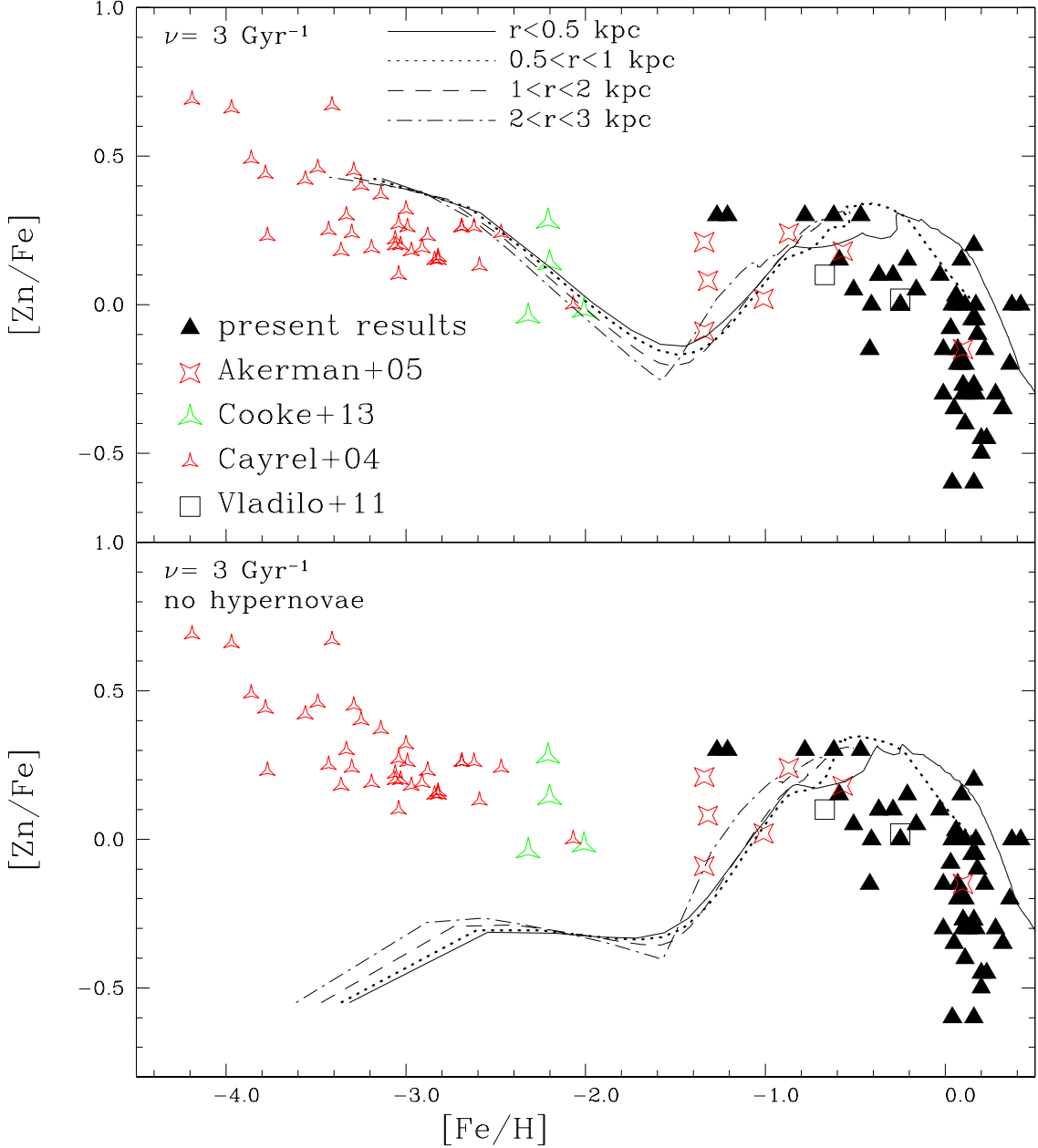,angle=0.,width=17.0 cm}
\caption{
Upper panel:
Comparison of the evolution of [Zn/Fe] with [Fe/H] in the stellar population
predicted by a chemical evolution model for a classical bulge
with star formation rate normalization $\nu_{\rm SF}=$ 3 Gyr$^{-1}$
and total mass of 1.5$\times$10$^{10}$ M$_{\odot}$,
with the [Zn/Fe] vs. [Fe/H] data for the present sample,
for the DLA samples by  Cooke et al. (2013), Akerman et al. (2005) and  Vladilo et al. (2011),
and for the halo stars from Cayrel et al. (2004).
The Fe abundances of Akerman et al. (2005) and  Vladilo et al. (2011)
have been corrected for dust depletion.
(No dust correction was needed for the Cooke et al. sample,
because their systems have [Fe/H] $<$ $-$2.)
Lower panel:
The same as the upper model, but using, at low metallicities,
the  yields of Woosley \& Weaver (1995) instead of those of hypernovae.
}
\label{amancioznfe} 
\end{figure*}

Since bulge stars are probes of bulge formation and evolution,
Fig. \ref{amancioznfe} shows the comparison of the
[Zn/Fe] vs. [Fe/H] in bulge stars as derived in the present paper
with a chemo-dynamical model describing a classical bulge.
The computed 
 models assume a specific star formation rate of $\nu_{\rm SF}=$ 3 Gyr$^{-1}$, a
baryonic mass of 2$\times$10$^9$ M$_{\odot}$,
and a dark halo mass $M_{H}$= 1.3$\times$10$^{10}$ M$_{\odot}$
(total mass of 1.5$\times$10$^{10}$ M$_{\odot}$,
thus reproducing the cosmological proportion
 of baryonic mass, $\Omega_b = 0.04$,
to total matter mass, $\Omega_m = 0.3$).
In the present calculations, we adopt
an $\Omega_m = 0.3$, $\Omega_{\Lambda}= 0.7$, $H_0=70$ km s$^{-1}$ Mpc$^{-1}$
cosmology, with the corresponding age of the universe of 13.47 Gyr.
(An age of 13.799$\pm$0.038 Gyr is the most recent and updated value from
the Planck satellite data as given by the
Planck collaboration: Adam et al. (2015).)

The evolution of the model is followed until 13 Gyr,
and it gives the evolution of the average Zn chemical abundance
of the stellar population for several radii.
As can be seen in Fig. \ref{amancioznfe}, its end point
falls in the locus of the data for the present-day stars 
in the bulge of the Galaxy.
The trend towards decreasing [Zn/Fe] ratio with increasing [Fe/H] 
for higher metallicities
is reproduced well by the model.
This is because although the bulge is formed rapidly
 in the classical scenario,
the star formation goes on for a few Gyr.
In the present case, the stellar mass is built up during
 at least $\approx 3$ Gyr,
which allows the contribution of type Ia supernovae (SNIa) to be relevant,
increasing the Fe abundance.
The ejecta of SNIa exhibit a very low [Zn/Fe] ratio.
The present model uses the SNIa yields of Iwamoto et al (1999),
[Zn/Fe] $\approx -1.2$ for a zero initial metallicity,
and [Zn/Fe] $\approx -1.6$ for a solar initial metallicity.
Therefore, as a result of the continuing star formation,
the classical bulge model predicts subsolar [Zn/Fe] ratios for higher 
metallicities,
as observed in the present sample of bulge stars.

The nucleo-synthetic nature of Zn is complex;
it is neither an $\alpha$-element nor a Fe peak element. 
Theoretical work on the nucleo-synthesis of Zn
often predicts that it may originate in massive stars, but
a complete network for production of Zn is not yet available. For instance, the classical
core-collapse SN II yield
calculations of Woosley and Weaver (1995) are known to underestimate the Zn 
abundance. One solution to this problem could be $ \alpha$-rich freeze-out neutrino 
winds, as predicted by Woosley and Hoffman (1992). On the other hand, 
Umeda \& Nomoto (2002)
 have produced nucleo-synthesis calculations in core-collapse explosions of 
massive  low metallicity stars that do show  large [Zn/Fe] for deeper mass cuts, 
smaller neutron excesses, and higher explosion energies. In the last case, the 
supernova would be classified as a hypernova
 as defined by Nomoto et al. (2013, and references therein).
 Therefore, in the nucleosynthesis input 
of our chemical calculations, we consider the 
core-collapse SN II models of Woosley and Weaver 
(1995), but, for lower metallicities, we use the results of  
high explosion-energy 
hypernovae (Umeda \& Nomoto 2002; 2003; 2005; Nomoto et al. 2006; 2013).

In summary, most of the bulge star data obtained in this work can be
 explained by a classical scenario of bulge formation.
The trend towards a decreasing [Zn/Fe] ratio with increasing [Fe/H]
seems to be reproduced well by the model.
In addition, the model also accounts for the abundances of the halo stars,
which can be thought of as a relic of
the same galaxy formation sequence of events that gave rise to the bulge.
The high [Zn/Fe] ratio of our calculations results from including 
hypernovae.
The sensitivity of our results to the nucleo-synthesis prescription is
shown in the lower panel of  \ref{amancioznfe},
in which the Woosley \& Weaver (1995) yields are used at low metallicities.
The resulting [Zn/Fe] is very low, which favours hypernovae
at low metallicities, as given in the upper panel. 
Finally, the decreasing trend of [Zn/Fe] 
at high metallicities is due to Fe enrichment from SNIa.

\subsection{Zn and alpha elements}

\begin{figure*}
\centering
\psfig{file=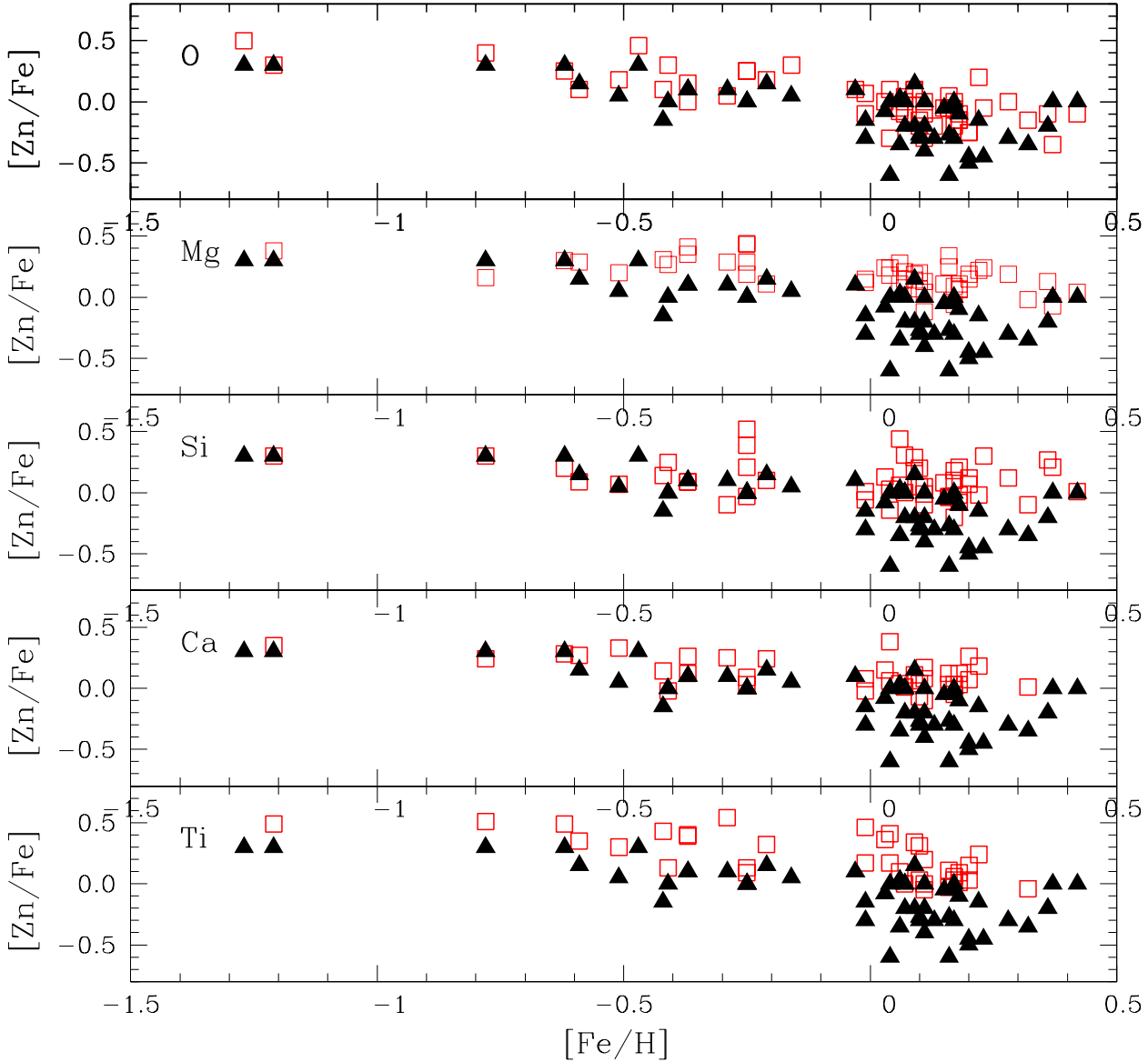,angle=0.,width=17.0 cm}
\caption{[Zn/Fe] vs. [Fe/H] for the present sample (filled black triangles),
compared with alpha-element abundances (open red squares), for
oxygen, magnesium, silicon, calcium, and titanium.
 For oxygen all values are the presently revised oxygen abundances.
}
\label{alpha} 
\end{figure*}

The Zn enhancement in metal-poor stars suggests that [Zn/Fe] behaves
like alpha elements. For this reason, in Fig. \ref{alpha} we compare
the Zn abundances with those for the alpha elements O, Mg, Si, Ca, and Ti,
derived in Lecureur et al. (2007), Zoccali et al. (2006),
and Gonzalez et al. (2011). For oxygen
the revised values given in Table \ref{cncorrected} for selected stars
 are plotted instead of the previous values.
The trend shown by Zn appears similar to that of the alpha elements
and, more strikingly, of oxygen and calcium. 
The low [Zn/Fe] for high metallicity
stars is compatible with the oxygen abundances.

\begin{figure}
\centering
\psfig{file=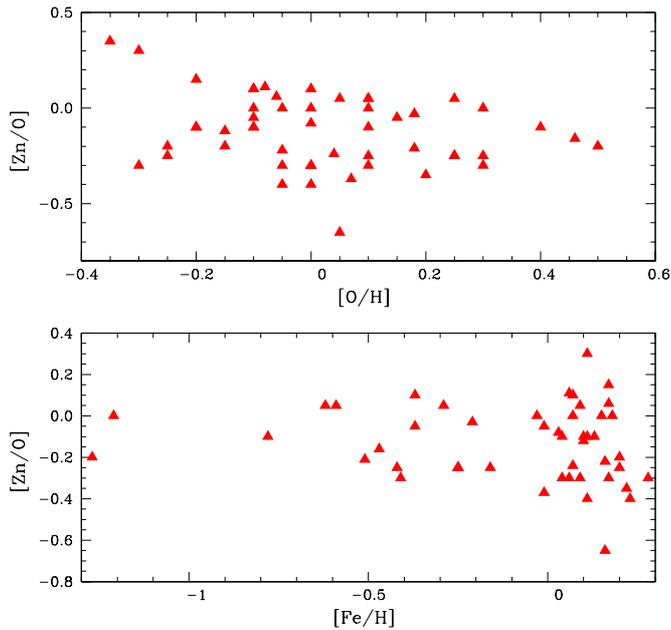,angle=0.,width=9.0 cm}
\caption{[Zn/O] vs. [Fe/H] and [Zn/O] vs. [O/H] for the present sample,
where the oxygen abundances are the revised ones from Table \ref{cncorrected}, otherwise those from Zoccali et al. (2006) and Lecureur et al. (2007).
}
\label{zno} 
\end{figure}

The variation in Zn in lockstep with alpha elements is made evident further
in Fig. \ref{zno}, where [Zn/O] shows essentially no trend with [Fe/H] or
with [O/H].
 
\section{Conclusions}

The iron-peak elements Sc, Mn, Cu, and Zn show a different chemical
enrichment pattern than do the even-Z iron-peak elements Fe and Ni.
In Barbuy et al. (2013), we confirmed that Mn behaves as a secondary
element with low [Mn/Fe] in metal-poor stars, by increasing with
increasing metallicity. In the present work we show that, in the metal-rich bulge stars, Zn-to-Fe
decreases with increasing metallicity, complementing
the long established
 high Zn abundances in metal-poor stars (e.g. Sneden et al. 1991;
Nissen \& Schuster 2011).

 Our main comments on the comparison between the data points and models
are the following: a) The most metal-rich bulge dwarfs from
 Bensby et al. (2011) show
a constant [Zn/Fe], which implies that
there is no contribution of SNe type I in the bulge, whereas  a decrease in
[Zn/Fe] in the present sample of giants, as derived here,
 implies that there is enrichment
from Type I, SNe as predicted by the models.  The bulge stars are
a complex mix of stellar populations of different ages and formation
processes.
b) The high [Zn/Fe] in very metal-poor stars favours enrichment from hypernovae,
as defined by Nomoto et al. (2013 and references therein)
acting at these low metallicities.  It is interesting to point out
that the hypernovae as defined by Nomoto et al. (2013 and
references therein) might be related to the spinstars as
defined by
Frischknecht et al. (2012) and Meynet et al. (2006) and discussed
in terms of early enrichment of the bulge in Chiappini et al. (2011).
  c) The drop in [Zn/Fe] for moderately
metal-poor stars ($-$2.2 $\simless$ [Fe/H] $\simless$ $-$1.6) corresponds to 
 the normal metal-poor supernovae, here using the  yields from Woosley
\& Weaver (1995).

For the DLA systems with measured
Fe abundances, it was crucial to correct for dust depletion. The Zn abundances
were also corrected for dust depletion, even if these corrections are smaller.
The chemical evolution models predict subsolar
[Zn/Fe] values at relatively high metallicities ([FeH]$\simgreat$-1.0),
as confirmed  for a few systems.

\begin{table*}
\begin{flushleft}
\small
\caption{Atmospheric parameters and C,N,O abundances
 adopted from Zoccali et al. (2006), Lecureur
et al. (2007), and Hill et al. (2011).
Column 11: revised C,N,O abundances for selected stars (see text), where
c indicates that the previous value is adopted, otherwise a new value
is reported. Columns 12, 13: [Zn/Fe] abundances with
Zn1 and Zn2 corresponding to the 4810 and 6362 {\rm \AA}
lines; Column 14: final mean Zn abundances.
 }             
\label{cncorrected}      
\centering          
\begin{tabular}{l@{}r@{}r@{}r@{}r@{}r@{}r@{}r@{}r@{}r@{}r@{}r@{}r@{}r@{}r@{}r@{}r@{}r@{}}     
\noalign{\smallskip}
\hline\hline    
\noalign{\smallskip}
\noalign{\vskip 0.1cm} 
Star & \phantom{-}\phantom{-}OGLE n$^{\circ}$ &  \phantom{-}\phantom{-}$V$ &
  $\phantom{-}$$\phantom{-}$T$_{\rm eff}$ & $\phantom{-}$log~$g$ &
 \phantom{-}\phantom{-}$\phantom{-}$v$_{\rm t}$
 & \phantom{-}\phantom{-}[Fe/H] & \phantom{-}\phantom{-}[C/Fe] & 
\phantom{-}\phantom{-}[N/Fe] & \phantom{-}\phantom{-}[O/Fe] & 
\phantom{-}\phantom{-}[C,N,O/Fe]$_{\rm corr}$ &
 \phantom{-}[Zn1/Fe] & \phantom{-}[Zn2/Fe] & \phantom{-}[Zn/Fe] & 
\phantom{-}Comments \\                            
 \noalign{\vskip 0.1cm}    
\noalign{\vskip 0.1cm}
\noalign{\hrule\vskip 0.1cm}
\noalign{\vskip 0.1cm}  
B6-b1& 29280c3 &16.14& 4400 & 1.8 & 1.6 & 0.07    &$-$0.16 & 0.39 & 0.04 
& c,c,c & 0.00    & $-$0.40  &  $-$0.20 & \\
B6-b2& 83500c6 &16.40&4200 &1.5   &1.4   &$-$0.01  & ---  & ---  & ---  
&  $-$0.2,0.6,$-$0.1  &  $-$0.15    &  ---  &  $-$0.15 & CN-strong  \\
B6-b3& 31220c2  &16.09& 4700 & 2.0 & 1.6 & 0.10    & $-$0.16& 0.11 & 0.19 
&  c,0.3,$-$0.15 & $-$0.25 &   $-$0.30 &   $-$0.27 &  \\  
B6-b4& 60208c7  &16.12& 4400 & 1.9 & 1.7 & $-$0.41 & $-$0.14& 0.53 & 0.53 
&  c,c,0.3 & 0.00    & 0.00      &    0.00 & \\
B6-b5& 31090c2  &16.09&4600 & 1.9 & 1.8  & $-$0.37 & $-$0.11& 0.56 & 0.33 
&  c,0.3,0.15    & +0.20   &  0.00   &  +0.10 &  \\
B6-b6**& 77743c7 &16.09&4600 & 1.9 & 1.8 & 0.11    & $-$0.03& 0.57 & 0.01 &
 $\leq$$-$0.1,0.7,0.0 & $-$0.30 &    $-$0.50    &  $-$0.40 &  \\
B6-b8& 108051c7  &16.29& 4100 & 1.6  & 1.3  & 0.03    &  0.08& 0.05 & 0.10 
&  0,c,0  &  $-$0.15 &  0.00  &   $-$0.08 & \\

B6-f1**& 23017c3 &15.96& 4200 & 1.6 & 1.5    & $-$0.01 &  0.05& 0.55 & 0.18 &
  0.0,0.35,0.07 & $-$0.30 &   ---    &  $-$0.30 & CN-strong \\
B6-f2& 90337c7 &15.91& 4700 & 1.7 & 1.5    & $-$0.51 & $-$0.04& 0.56 & 0.39 
&  0,0.05,0.18 & +0.05   &  +0.05 &   +0.05 &  \\
B6-f3& 21259c2 &15.71& 4800 & 1.9 & 1.3 & $-$0.29 & $-$0.09& 0.53 & 0.18 
&  c,0.3,0.05   & +0.20   &  0.00 &  +0.10 & \\
B6-f5**& 33058c2 &15.90& 4500 & 1.8 & 1.4 & $-$0.37 &  0.37& 0.53 &---&
  $-$0.1,0.15,0.0  & 0.00   &  +0.2  &  +0.10 &\\
B6-f7** & 100047c6 &15.95& 4300 & 1.7& 1.6  & $-$0.42 &  0.42& 0.57 &---&
 0,0.3,0.1 & $-$0.30  & \phantom{-}0.00  &  $-$0.15 & \\
B6-f8 & 11653c3 &15.65&4900 & 1.8 & 1.6    & 0.04    & $-$0.11& 0.51 &$-$0.17 & 
 c,0.35,$-$0.3 & $-$0.60 &  ---  &   $-$0.60 & CN-strong \\

BW-b2 & 214192 &16.58& 4300 & 1.9 & 1.5  & 0.22    &  0.05& 0.25 & 0.23 & 
 $-$0.1,0.1,$-$0.1 & $-$0.30 &  0.00 & $-$0.15 & \\
BW-b4& 545277 &16.95& 4300 & 1.4  & 1.4   & 0.07    &  --- & ---  &  --- 
&$-$0.1,0,$-$0.1 & --- & 0.00 &  0.00 & \\
BW-b5& 82760 &16.64& 4000 & 1.6  & 1.2   & 0.17    & 0.06 & 0.56 & 0.09 
& c,0,0 &  $-$0.30 &  --- &   $-$0.30  & CN-strong \\
BW-b6& 392931 &16.42& 4200 & 1.7 & 1.3  & $-$0.25 & 0.05 & 0.56 & 0.09
 &$-$0.3,0.9,0.25 & 0.00 & 0.00 &  0.00  & \\
BW-b7 & 554694 &16.69& 4200 & 1.4 & 1.2   & 0.10    &  --- & ---  &  --- 
&$-$0.23,$-$0.1,$-$0.2 & $-$0.30 & $-$0.30 & $-$0.30&\\

BW-f1 & 433669 &16.14& 4400 & 1.8 & 1.6  & 0.32    & $-$0.26& 0.24 &$-$0.02 &
 $-$0.2,0.5,$-$0.15 &  $-$0.35 &  ---  &   $-$0.35 & CN-strong \\
BW-f4& 537070 &16.07&4800 & 1.9  & 1.7   & $-$1.21 &  0.04& 0.54 &---
&c,c,0.3 &  +0.30  & +0.30   &  +0.30 &  \\
BW-f5**& 240260 &15.88& 4800 & 1.9 & 1.3    & $-$0.59 &  0.03& 0.53 & 0.31 &
 c,0.2,0.1 &  +0.30  &  0.00 &  +0.15 & \\
BW-f6& 392918  &16.37& 4100 & 1.7 & 1.5 & $-$0.21 &  0.08& 0.58 & 0.46 
& c,0.4,0.18 &  0.00   & +0.30    &  +0.15 & \\
BW-f7 & 357480 &16.31& 4400 & 1.9 & 1.7 & 0.11 & $-$0.10& 0.30 &0.03 & 
 $-$0.2,0.6,$-$0.1 & $-$0.20 &  --- &  $-$0.20 & CN-strong \\
BW-f8 & 244598 &16.00& 5000 & 2.2 & 1.8  & $-$1.27 & 0.03 & 0.53 &  --- 
& 0,+0.2,+0.5 & +0.30   & +0.30  &  +0.30 &  \\

BL-1& 1458c3 & 15.37 & 4500 & 2.1  & 1.5  & $-$0.16 &  0.03 & 0.18 & 0.26 
&  0.15,0.4,0.3 &   +0.10  & 0.00  &  +0.05 & \\
BL-3& 1859c2  & 15.53 & 4500 & 2.3 & 1.4  & $-$0.03 & $-$0.07 & 0.18 & 0.22 
&c,0.3,0.1 &  +0.20 & 0.00  &  +0.10 &  \\
BL-4**& 3328c6 & 14.98 & 4700 & 2.0  & 1.5  & 0.13    & $-$0.04 & 0.41 & 0.02 &
 $-$0.1,0.3,$-$0.2 & $-$0.30 &  $-$0.30      &  $-$0.30 & \\
BL-5**& 1932c2  & 15.39 & 4500 & 2.1 & 1.6    & 0.16 &  0.04 & 0.33 & 0.07 &
 $-$0.02,c,$-$0.05 & $-$0.25 &  $-$0.30   &  $-$0.27 & \\
BL-7& 6336c7  & 15.33 & 4700 & 2.4 & 1.4   & $-$0.47 & $-$0.17 & 0.38 & 0.46 
& c,c,0.3$^t$ & +0.30   &  +0.30   &  +0.30   & telluric \\ 

B3-b1 & 132160C4 &16.35& 4300 & 1.7 & 1.5   & $-$0.78 &$-$0.10 & 0.45 & 0.55 
& 0.1,0.6,0.4 & ---  & +0.30  &  +0.30 & \\
B3-b2& 262018C7  &16.63& 4500 & 2.0 & 1.5    & 0.18    &$-$0.13 & 0.42 & 0.12 
& c,0,$-$0.1 &  $-$0.30 &  +0.10  &   $-$0.10 &   \\
B3-b3**& 90065C3 &16.59& 4400 & 2.0  & 1.5 & 0.18 &$-$0.09 & 0.46 &$-$0.19 &
 0,0,$-$0.15 & ---     &   --- &   --- & CN-strong \\
B3-b4& 215681C6 &16.36& 4500 & 2.1 & 1.7    & 0.17    &$-$0.16 & 0.39 &$-$0.06 
&c,c,c & 0.00  &  ---  &   0.00 &  \\
B3-b5& 286252C7 &16.23& 4600 & 2.0 & 1.5    & 0.11    &$-$0.15 & 0.40 & 0.00 
& $-$0.2,0,$-$0.3 & 0.00   & ---  & 0.00  & CN-strong  \\
B3-b7& 282804C7 &16.36& 4400 & 1.9 & 1.3     & 0.20    &$-$0.16 & 0.39 &$-$0.06 
&c,0,$-$0.25 & $-$0.50 &  --- &  $-$0.50 & CN-strong \\
B3-b8& 240083C6 &16.49& 4400 & 1.8 & 1.4     & $-$0.62 &$-$0.16 & 0.39 & 0.52 
&c,0.3,0.25 & +0.30   & +0.30  &  +0.30 & \\
B3-f1**& 129499C4 &16.32& 4500 & 1.9 & 1.6 & 0.04    & 0.09 & 0.44 & 0.19 & 
 0,0.4,0.1 & 0.00 &  0.00 &  0.00 &  \\
B3-f2& 259922C7 &16.54& 4600 & 1.9 & 1.8    & $-$0.25 &$-$0.15 & 0.40 &---
& c,c,(0.2) & 0.00    & 0.00 & 0.00 & telluric \\
B3-f3**& 95424C3 &16.32& 4400 & 1.9 & 1.7 & 0.06 &$-$0.08 & 0.47 & $-$0.08&
 0,0,c &  +0.2 &  $-$0.15 &  0.03 &  \\
B3-f4& 208959C6 &16.51& 4400 & 2.1 & 1.5    & 0.09    & 0.10 & 0.45 & 0.43 
& c,0.1,0.1 & 0.00 & +0.30 & +0.15 & \\
B3-f5& 49289C2  &16.61& 4200 & 2.0 & 1.8    & 0.16    &$-$0.06 & 0.49 & 0.09 
&c,c,$-$0.05 & ---  & --- &  --- & CN-strong \\
B3-f7& 279577C7 &16.28&  4800 & 2.1 & 1.7    & 0.16    &$-$0.02 & 0.33 & 0.05 
& c,c,c & $-$0.60 & $-$0.60  &  $-$0.60 & \\
B3-f8& 193190C5 &16.26& 4800 & 1.9 & 1.5  & 0.20    &$-$0.17 & 0.38 & 0.00 
&c,0.28,$-$0.25 & $-$0.30 &  $-$0.60 &  $-$0.45 & \\
BWc-1**& 393125 &16.84& 4476 & 2.1 & 1.5    & 0.09    & 0.12 & 0.50 & 0.19 &
 0.05,0.3,0.1 &  $-$0.30 &  $-$0.10 &  $-$0.20 & \\
BWc-2 & 545749 &17.19& 4558 & 2.2 & 1.2   & 0.18    &$-$0.09 & 0.52 & 0.07 
& $-$0.2,0.2,$-$0.15 & ---     &  ---  &  ---  & CN-strong  \\  
BWc-3& 564840 &16.91& 4513 &2.1  &1.3   & 0.28    & 0.04 & 0.51 & 0.19 
& $-$0.1,0.4,0 & $-$0.30 &   ---  &  $-$0.30  & CN-strong \\
BWc-4 & 564857 &16.76& 4866 & 2.2& 1.3 & 0.06    & 0.36 & 0.28 & 0.05 &
 $-$0.1,0.05,$-$0.05 & $-$0.30  & $-$0.40 &  $-$0.35  & \\
BWc-5** & 575542 &16.98& 4535 &2.1  & 1.5 &0.42  &$-$0.01& 0.59 & 0.04 &
 $-$0.05,0.4,$-$0.1 & \phantom{+}0.00  &  --- &  \phantom{+}0.00  & CN-strong \\
BWc-6& 575585 &16.74& 4769 &2.2 &1.3 &$-$0.25  &$-$0.20 & 0.69 & 0.43 
& c,c,0.25 &   \phantom{+}0.00    &  ---  &    \phantom{+}0.00 & CN-strong \\  
BWc-7 & 67577 & 17.01 & 4590 &2.2 &1.1 &$-$0.25  &$-$0.20 & 0.50 & 0.44 
&c,0.3,0.25 &   \phantom{+}0.00 & \phantom{+}0.00 & \phantom{+}0.00 &   \\   
BWc-8& 78255 & 16.97 & 4610 &2.2 &1.3  & 0.37    &$-$0.22 & 0.47 &$-$0.07 
& c,0.1,$-$0.35 & \phantom{+}0.00    &  ---  & \phantom{+}0.00  & CN-strong \\  
BWc-9**& 78271 & 16.90 & 4539 &2.1 &1.5  & 0.15    &$-$0.13 & 0.77 & 0.11 &
 $-$0.1,0.2,$-$0.05 & 0.00  &  $-$0.10 &  $-$0.05    & \\   
BWc-10& 89589 & 16.70  & 4793 &2.2& 1.3   & 0.07   &$-$0.15 & 0.54 & 0.15 
&  $-$0.2,0.3,0 &  \phantom{+}0.00    &  \phantom{+}0.00 & \phantom{+}0.00  & \\   
BWc-11**& 89735 &  16.69  & 4576 &2.1 & 1.0  & 0.17   &$-$0.14 & 0.56 &---
& $-$0.2,0,$-$0.2 &  $-$0.10  &  \phantom{+}0.00 &   $-$0.05  &  \\   
BWc-12**& 89832 & 16.92  & 4547 &2.1&1.3  & 0.23   & 0.19 & 0.29 & 0.24 &
 $-$0.1,0.1,$-$0.05 & $-$0.30 &   $-$0.60 &  $-$0.45  & \\    
BWc-13**& 89848 & 16.73  & 4584 &2.1 &1.1 & 0.36 & 0.12 & 0.51 & 0.13 &
 0,$-$0.15,$-$0.1 &  $-$0.10 &  $-$0.30 &  $-$0.20 &  \\  
\noalign{\vskip 0.1cm}
\noalign{\hrule\vskip 0.1cm}
\noalign{\vskip 0.1cm}  
\hline                  
\end{tabular}
\end{flushleft}
\end{table*}  

\begin{acknowledgements}
      BB acknowledges partial financial support by CNPq, CAPES, and FAPESP.
CRS acknowledges a CAPES/PROEX PhD fellowship.
MZ and DM acknowledge
 support by the Ministry of Economy, Development, and Tourism's
Millenium Science Initiative through grant IC120009, awarded to
The Millenium Institute of Astrophysics, MAS, and
from the  BASAL Center for Astrophysics and Associated
 Technologies PFB-06 and FONDECYT Projects 1130196 and 1150345.
SO acknowledges the Italian Ministero dell'Universit\`a e della Ricerca
Scientifica e Tecnologica (MURST), Italy. 
\end{acknowledgements}

\end{document}